\documentstyle[12pt,psfig,epsf]{article}
\newcommand{\be}{\begin{eqnarray}}
\newcommand{\ee}{\end{eqnarray}}
\newcommand{\beno}{ \begin{eqnarray*}}
\newcommand{ \eeno}{\end{eqnarray*}}
\newcommand{\raf}[1]{(\ref{#1})}
\newcommand{\dsl}{\partial \!\!\! /}
\newcommand{\gmf}{\gamma_5}
\newcommand{\dsig}{(\delta \sigma)}
\newcommand{\dpi}{(\delta \pi)}
\begin{document}

\bibliographystyle{/homeb/vkoch/tex/bst/try}
 \begin{titlepage}
\hspace{11cm}
{\large LBNL--39463}

\hspace{11cm} 
{\large UC--413}

\vspace{.7cm}
\begin{center}
\ \\
{\large {\bf Aspects of  Chiral Symmetry}}\footnote{This work was supported
 by the Director, 
Office of Energy Research, Office of High Energy and Nuclear Physics, 
Division of Nuclear Physics, Division of Nuclear Sciences, of the U.S. 
Department of Energy under Contract No. DE-AC03-76SF00098.}
\vspace{2cm}
\ \\
{\large Volker Koch}
\ \\
\ \\
{\it Nuclear Science Division, Lawrence Berkeley National Laboratory,\\
University of California,\\
Berkeley, CA, 94720, U.S.A.}\\
\ \\
\today \\

\vspace{2cm}
{\large \bf Abstract}\\
\vspace{0.2cm}
\end{center}
\begin{quotation}
This article is an attempt to a pedagogical 
introduction and review into the elementary concepts of 
chiral symmetry in nuclear physics. Effective chiral models such as the linear
and nonlinear sigma model will be discussed as well as the essential ideas of
chiral perturbation theory. Some applications to the physics of 
ultrarelativistic heavy ion collisions will be presented.
\end{quotation}
\end{titlepage}
\section{Introduction}
Chiral symmetry is a symmetry of QCD in the limit of vanishing quark masses.
We know, however, that the current quark masses are finite. But compared with
hadronic scales the masses of the two lightest quarks, up and down,
are very small, so that chiral symmetry may be considered an
approximate symmetry of the strong interactions. 

Long before QCD was believed to be the theory of strong interactions, 
phenomenological indications for the existence of chiral symmetry came
from the study of the nuclear beta decay. There one finds, that the weak
coupling constants for the vector and axial-vector hadronic-currents, 
$C_V$ and $C_A$, did not (in case of $C_V$) or only slightly 
($25 \%$ in case of $C_A$) differ from those for the leptonic counterparts.
Consequently strong interaction `radiative' corrections to the weak 
vector and axial vector `charge' are absent. 
The same is true for the more familiar case of the electric charge, and there
we know that it is its conservation, which protects it from radiative
corrections. Analogously, we expect the weak vector and axial vector charge, or
more generally, currents, to be conserved  due to some symmetry of the strong
interaction. In case of the vector current, the underlying symmetry is the
well known isospin symmetry of the strong interactions and thus the hadronic
vector current is identified with the isospin current. The identification of
the axial current, on the other hand is not so straightforward. This is due to
another, very important and interesting feature of the strong interaction,
namely that the symmetry associated with the conserved axial vector current
is `spontaneously broken'. By that, one means that while the Hamiltonian
possesses the symmetry, its ground state does not. An important consequence of
the spontaneous breakdown of a symmetry is the existence of a massless mode,
the so called Goldstone-boson. In our case, the Goldstone boson is the pion. If
chiral symmetry were a perfect symmetry of QCD, the pion should be
massless. Since chiral symmetry is only approximate, we expect the
pion to have a finite but small (compared to all other hadrons) mass. This is
indeed the case!

The fact that the pion is a Goldstone boson is of great practical
importance. Low energy/temperature hadronic processes are 
dominated by pions and thus all observables can be expressed as an expansion 
in pion masses and momenta. This is the basic idea of chiral perturbation
theory, which is very successful in describing threshold pion physics.

At high temperatures and/or densities one expects to `restore' chiral 
symmetry. By that one means, that, unlike the ground state, the state at high
temperature/density posses the same symmetry as the Hamiltonian (the symmetry
of the Hamiltonian of course will not be changed). As a consequence of this so
called `chiral restoration' we expect the absence of any Goldstone modes and
thus the pions, if still present, should become as massive as all other
hadrons\footnote{If of course chiral restoration and deconfinement take place
at the same temperature, as current lattice gauge calculations suggest, the
concept of hadrons in the restored phase may become meaningless.}.
To create a system of restored chiral symmetry in the laboratory  
is one of the major goals of the ultra-relativistic heavy ion experiments.

This article is intended to serve as an  introduction into the ideas
of chiral symmetry in particular for experimentalists interested or working in
this field. Thus emphasis will be put on the ideas and concepts rather than 
formalism. Consequently, most arguments presented will be heuristic and/or 
based on simple effective models. References will be provided for those seeking
more rigorous derivations.

In the first section we will introduce some basic concepts of quantum field
theory, which are necessary to discuss the effect of symmetries on the
dynamics. Then we will introduce the chiral symmetry transformations and
derive some results, such as the Goldberger-Treiman relation.
In the second section we will present the linear sigma model as the most
simple 
effective chiral model. Using this rather intuitive model we will discuss
explicit chiral symmetry breaking. As an application we will consider
pion-nucleon scattering.
The third section will be devoted to the so called nonlinear sigma model, which
then serves as a basis for the introduction into chiral perturbation theory.
In the last  section we will give some examples for chiral symmetry in
the physics of hot and dense matter.

When picking reference, I gave preference to textbooks and review articles over
the original work. I felt this provides a better basis for those who want to
study the subject more thoroughly than it is presented here.
 If not stated otherwise, natural units, i.e. $\hbar = c = 1$
 and the conventions of Bjorken and
Drell \cite{BD65} for metric, gamma-matrices etc are used.

\section{Theory Primer}
\newcommand{\qdot}{\dot{q}}
\subsection{Basics of quantum field theory}
In this section we briefly review the essentials of quantum field theory, which
are required for the understanding of chiral symmetry . For a detailed and
thorough exposition of this subject we refere the reader to the standard
textbooks such as the one by Bjorken and Drell \cite{BD65} and Ramond
\cite{Ram89}.

A field theory is usually written down in the Lagrangian formulation. 
Let's start out with what we
know from classical mechanics of a point particle. 
There, one obtains the equations of motion from
the Hamilton principle, where one requires that the variation of the action
$ S = \int_{t_1}^{t_2} dt \, L(q,\qdot,t) $ vanishes
\be
\delta S = 0 \,\, \Rightarrow \frac{d}{dt} \frac{\partial L}{d \qdot} - 
\frac{\partial L}{\partial q} = 0
\ee
Here $S$ is called the action and $L = T-V$ is the Lagrange-function.
For example, Newtons equations of motion for a particle in a potential $V(q)$
derive from
\be
L = \frac12 m \qdot^2 - V(q)&&\\
\Rightarrow m \ddot{q} + \frac{\partial V}{\partial q} = 0 &\Leftrightarrow&
m \ddot{q} = - \frac{\partial V}{\partial q} = F
\ee
If one goes over to a field theory, the coordinates $q$ are replaced by the
fields $\Phi(x,t)$ and the velocities $\qdot$ are replaced by the derivatives
of the fields. Furthermore, one requires that the fields and their derivatives
vanish at infinity
\be
q &\rightarrow& \Phi(x,t)\\
\qdot &\rightarrow& \partial_\mu \Phi(x,t) \equiv 
\frac{\partial \Phi(x,t)}{\partial x^\mu}
\ee
The Lagrange-function is then given by the spatial integral over the Lagrangian
density, ${\cal L}$, or Lagrangian, as we shall call it from now on
\be
L & = & \int d^3 x \, {\cal L}(\Phi(x,t),\partial_\mu \Phi(x,t),t)\\
S & = & \int_{t_1}^{t_2} dt \, L   = \int d^4 x \, 
{\cal L}(\Phi(x,t),\partial_\mu \Phi(x,t),t )
\ee
Lorentz invariance implies that the action $S$ and thus the 
Lagrangian ${\cal L}$ transform like Lorentz-scalars.
The equations of motion for the fields are again obtained by requiring that the
variation of the action S vanishes. This variation is carried out by a
variation of the fields\footnote{From now on, we will denote the space-time
dependence of the field explicitly only when this is essential for the
understanding.}
\be
\Phi &\rightarrow& \Phi + \delta \Phi\\
\partial_\mu \Phi &\rightarrow&  \partial_\mu \Phi + \delta(\partial_\mu \Phi )
\ee
with
\be
\delta(\partial_\mu \Phi ) = \partial_\mu(\Phi + \delta \Phi) - \partial_\mu
 \Phi =
\partial_\mu( \delta \Phi )
\label{eq.1.3}
\ee
In addition, as in classical mechanics, one requires that the variation at the 
boundaries vanishes
\be
\delta \Phi (t_1) = \delta \Phi (t_2) = 0, \, \, \, \rm etc.
\ee
Consequently,
\be
\delta S & = & 
\int_{t_1}^{t_2} dt \int d^3 x \, \, 
{\cal L}(\Phi + \delta \Phi, \partial_\mu \Phi 
+ \delta (\partial_\mu \Phi ) )  
- {\cal L}(\Phi , \partial_\mu \Phi  )  
\nonumber \\ 
& = & \int_{t_1}^{t_2} dt \int d^3 x \, 
\left[ {\cal L}(\Phi , \partial_\mu \Phi  ) + 
\frac{\partial {\cal L}}{\partial \Phi} \delta \Phi + 
\frac{\partial {\cal L}}{\partial (\partial_\mu \Phi )} 
\delta (\partial_\mu \Phi ) \right] - {\cal L}(\Phi , \partial_\mu
 \Phi )
\nonumber \\ 
& = & \int_{t_1}^{t_2} dt \int d^3 x \, \left( \frac{\partial
{\cal L}}{\partial \Phi} \delta \Phi + \frac{\partial {\cal L}}{\partial
(\partial_\mu \Phi)} \partial_\mu (\delta \Phi ) \right)
\label{eq.1.4}
\ee
where eq. \raf{eq.1.3} has been used. 
The derivatives of ${\cal L}$ with respect
to the fields are so called functional derivatives, but for all practical
purposes they just work like `normal' derivatives, where ${\cal L}$ is
considered a function of the fields $\Phi$. 
Partial integration\footnote{The `surface terms' do not contribute, because we
have required that the fields and their variation vanish at the boundary.} 
of the second term of eq. \raf{eq.1.4} finally gives
\be
0 = \delta S  = \int_{t_1}^{t_2} dt \int d^3 x \left( 
\frac{\partial {\cal L}}{\partial \Phi}  - 
\partial_\mu ( \frac{\partial {\cal L}}{\partial (\partial_\mu \Phi)}) \right)
(\delta \Phi )
\ee
Since the variation $\delta
\Phi$ are arbitrary, we obtain the following equations of motion

\be
\frac{\partial {\cal L}}{\partial \Phi}  - 
\partial_\mu ( \frac{\partial {\cal L}}{\partial (\partial_\mu \Phi)})  = 0
\label{eq.1.15}
\ee

If we are dealing with more than one field, such as in case of pions, where we
have three different charge states, the equations of motion have the same
form as in eq. \raf{eq.1.15} only that the fields carry now an 
additional index labeling the different fields under consideration
\be
\frac{\partial {\cal L}}{\partial \Phi_i}  - 
\partial_\mu ( \frac{\partial {\cal L}}{\partial (\partial_\mu \Phi_i)})  = 0
\label{e.o.m.}
\ee

So far, we have only dealt with a classical field theory. The fields are
quantized by requiring equal time 
canonical commutation relations between the fields 
$\Phi(x)$ and
their canonical momenta $\Pi(x)$, which are given by\footnote{For details see
\protect\cite{BD65}. A nice discussion on how the quantization rules of a
contiuum field theory relate to those of the quatum mechanics of a point
particle can be found in ref. \cite{HT62}}
\be
\Pi(x) = \frac{\partial {\cal L}}{\partial (\partial \Phi(x) / \partial t)}
\ee
\be
[\Phi(x,t), \Phi (x',t)] &=& 0 
\nonumber \\
\mbox{} [\Pi (x,t), \Pi (x',t)] &=& 0 
\nonumber \\  
\mbox{} [\Pi(x,t), \Phi(x',t)] &=&  -i \delta^3(x - x')
\ee
In case of fermions, the commutators have to be replaced by anti-commutators
due to the antisymmetrization properties (Pauli-principle) of the fermions.
As a result of the quantization, the fields are now Hilbert-space operators
acting on a given quantum state $| \phi >$.

\noindent
As an example let us consider the Lagrangian of a free boson and fermion field
respectively. \\

\noindent
{\bf (i)} free scalar bosons of mass $m$:
\be
{\cal L_{K.G.}} &=&\frac12 (\partial_\mu \Phi \partial^\mu \Phi) - \frac12 m^2
\Phi^2 
\label{L_KG}
\\
&\Rightarrow& \frac{\partial {\cal L}}{\partial \Phi} = - m^2 \Phi
\\
&\Rightarrow& \partial_\mu ( \frac{\partial {\cal L}}{\partial (\partial_\mu
\Phi_i)}) = \partial_\mu \partial^\mu \Phi 
\ee
Thus, according to eq. \raf{e.o.m.} the equation of motion is 
\be 
(\partial_\mu \partial^\mu  + m^2) \Phi = 
(\partial_t^2 - \nabla^2 +m^2 )\Phi = 0 
\ee
which is just the well known Klein-Gordon equation for a free boson.\\
\ \\
{\bf (ii)} free fermions of mass $m$ 
\be
{\cal L_{F.D.}} = \bar{\psi}(i \gamma_\mu \partial^\mu - m) \psi 
\label{l.f.d.}
\ee
By using the conjugate field $\bar{\psi}$  in the equation of motion 
\raf{e.o.m.}
\be
&\Rightarrow& \frac{\partial {\cal L}}{\partial \bar{\psi}} = (i \gamma_\mu
\partial^\mu - m) \psi
\\
&\Rightarrow& \frac{\partial {\cal L}}{\partial (\partial_\mu \bar{\psi})} = 0
\ee
we obtain the Dirac equation for $\psi$ :
\be
(i \gamma^\mu \partial_\mu - m) \psi = 0
\ee
whereas inserting $\psi$ for $\Phi_i$ in \raf{e.o.m.} leads to the conjugate 
Dirac equation
\be
\bar{\psi} (i \gamma^\mu \stackrel{\leftarrow}{\partial}_\mu + m)  = 0
\ee

\subsection{Symmetries}
\label{symmetries}

One of the big  advantages of the Lagrangian  formulation
is that symmetries of the Lagrangian lead to conserved quantities (currents).
In classical mechanics we know  that symmetries of the Lagrange
function imply conserved quantities. For example, if the Lagrange function is
independent of space and time, momentum and energy are conserved, respectively.

Let us assume that {\cal L} is symmetric under a transformation of the fields
\be
\Phi \longrightarrow \Phi + \delta \Phi
\ee
meaning
\be
{\cal L}(\Phi + \delta \Phi) &=& {\cal L}(\Phi)
\\
\Rightarrow 0 = {\cal L}(\Phi + \delta \Phi) - {\cal L}(\Phi) &=&
\frac{\partial{\cal L}}{\partial \Phi} \delta \Phi +  
\frac{\partial {\cal L}}{\partial (\partial_\mu \Phi)} \delta(\partial_\mu
\Phi)
\ee
where we have expanded the first term to leading order in $\delta \Phi$.
Using eq. \raf{eq.1.3} and the equation of motion \raf{eq.1.15} we have
\be
0 &=& \left( \partial_\mu \frac{\partial{\cal L}}{\partial \Phi}\right) 
\delta \Phi +
 \frac{\partial {\cal L}}{\partial (\partial_\mu \Phi)} \left( 
\partial_\mu \delta  \Phi \right)
\nonumber \\ 
&=& \partial_\mu \left( \frac{\partial {\cal L}}{\partial (\partial_\mu \Phi)}
\delta  \Phi \right) 
\ee
so that
\be
J_\mu = \frac{\partial {\cal L}}{\partial (\partial_\mu \Phi_i)} \delta  \Phi_i
\label{eq.1.50}
\ee
is a conserved current, with $\partial^\mu J_\mu = 0$. 
In the last equation we have included the indices for
possible different fields $\Phi_i$. 

As an example, let us discuss the case of a unitary transformation on the
fields, such as e.g. an isospin rotation among pions. For obvious reasons
unitary transformations are the most common ones,  and the chiral
symmetry transformations also belong to this class.
\be
\Phi_i \longrightarrow \Phi_i - i \Theta^a T^a_{ij} \Phi_j
\label{trans}
\ee
where $\Theta^a$ corresponds to the rotation angle and $T^a_{ij}$ is a matrix,
usually called the generator of the transformation (isospin matrix in case of
isospin rotations). The index $a$ indicates that there might be several
generators associated with the symmetry transformation (in case of isospin
rotations, we have three isospin matrices). 
Equation \raf{trans} corresponds to the expansion for small angles of the
general transformation
\be
\vec{\Phi} \longrightarrow e^{-i \Theta^a \hat{T}^a } \vec{\Phi}
\ee
where the vector on $\vec{\Phi}$ indicates  the several components of the field
$\Phi$ such as $\pi^+$, $\pi^-$ and $\pi^0$. 
From eq. \raf{eq.1.50} and eq. \raf{trans} we 
find the following expression for the conserved currents
\be
J_\mu^a = -i \frac{\partial {\cal L}}{\partial (\partial_\mu \Phi_j)} \,
T^a_{jk} \Phi_k 
\label{conserved}
\ee
where we have divided by the angle $\Theta^a$.
This current is often referred to as a Noether current, after E. Noether who
first showed its existence\footnote{Note, that some of the Noether
currents are not conserved on the  quantum-level. 
In other words, not every symmetry of the classical field
theory has a quantum analog.  If this is not the case one speaks of anomalies.
For a discussion of anomalies, see e.g. \cite{DGH92}.}.

Of course, a conserved current leads to a conserved charge
\be
Q = \int d^3 x J_0(x); \,\,\,\,\,   \frac{d}{d \, t} Q = 0
\ee

Finally, let us add a small symmetry breaking term to the Lagrangian
\be
{\cal L} = {\cal L}_0 + {\cal L}_1
\ee
where ${\cal L}_0$  is symmetric with respect to a given symmetry
transformation of the fields and ${\cal L}_1$ breaks this
symmetry. Consequently, the variation of the Lagrangian ${\cal L}$ does not
vanish as before but is given by
\be
\delta {\cal L} = \delta {\cal L}_1
\ee
Following the steps above, we can easily convince ourselves, that the variation
of the Lagrangian can still be expressed as the divergence of a current, which
is given by eq. \raf{eq.1.50} or \raf{conserved}, in case of unitary
transformations of the fields. Thus we have
\be
\delta {\cal L} = \delta {\cal L}_1 = \partial^\mu J_\mu
\label{unconserved}
\ee
Since $\delta {\cal L}_1 \neq 0$ the current $J_\mu$ is not conserved. Relation
\raf{unconserved} nicely shows how the symmetry breaking term of the Lagrangian
is related to the non-conservation of the current. It will also prove very
useful when we later on introduce the slight breaking of chiral symmetry due to
the finite quark masses.

\subsubsection{Example: Massless fermions}
\label{massless}

As an example for the Noether current, let us consider the Lagrangian of two
flavors of massless fermions. Since we will only discuss  transformations on 
the fermions, the results will be directly applicable to massless QCD.

The Lagrangian is given by (see eq. \raf{l.f.d.})

\be
{\cal L} = i \bar{\psi}_j \dsl \psi_j
\ee
where the index `$j$` refers to the two different flavors, let's say `up' and
`down', and $\dsl$ is the usual shorthand for $\partial_\mu \gamma^\mu$.\\

\noindent
{\bf (i)} Consider the following transformation
\be
\Lambda_V: \,\, \psi \longrightarrow e^{-i \frac{\vec{\tau}}{2} \vec{\Theta } }
\psi \, \simeq (1 - i \frac{\vec{\tau}}{2} \vec{\Theta} ) \psi
\label{vec_1}
\ee
where $\vec{\tau}$ refers to the Pauli - (iso)spin- matrices, and where we
have switched to a iso-spinor notation for the fermions, $\psi = (u,d)$. The
conjugate field, $\bar{\psi}$ transforms under $\Lambda_V$ as follows
\be
\bar{\psi} \, \longrightarrow e^{ + i \frac{\vec{\tau}}{2} \vec{\Theta} }
 \bar{\psi} \simeq
(1 + i \frac{\vec{\tau}}{2} \vec{\Theta} ) \bar{\psi}
\label{vec_2}
\ee
and, hence, the Lagrangian is invariant under $\Lambda_V$
\be
i \bar{\psi} \dsl \psi  &\longrightarrow& 
i \bar{\psi} \dsl \psi - i \vec{\Theta} \left( \bar{\psi} i \dsl 
\frac{\vec{\tau}}{2} \psi -  \bar{\psi}  \frac{\vec{\tau}}{2} i \dsl \psi 
\right)
\nonumber \\ 
& = &   i \bar{\psi} \dsl \psi
\ee
Following eq. \raf{conserved} the associated conserved current is
\be
V_\mu^a = \bar{\psi} \, \gamma_\mu \frac{\tau^a}{2} \,  \psi
\label{vec_current}
\ee
and is often referred to as the `vector-current'.\\

\noindent
{\bf (ii)} Next consider the transformation
\be
\Lambda_A: &\,\,& \psi \longrightarrow 
e^{-i \gmf \frac{\vec{\tau}}{2} \vec{\Theta } }
\psi \, = (1 - i \gmf \frac{\vec{\tau}}{2} \vec{\Theta} ) \psi
\label{ax_1}
\\
&\Rightarrow& \bar{\psi} \, \longrightarrow 
e^{ - i \gmf \frac{\vec{\tau}}{2} \vec{\Theta} }
 \bar{\psi} \simeq
(1 - i \gmf \frac{\vec{\tau}}{2} \vec{\Theta} ) \bar{\psi}
\label{ax_2}
\ee
where we have made use of the anti-commutation relations of the gamma matrices,
specifically, $\gamma_0 \gmf = - \gmf \gamma_0$. The Lagrangian transforms
under  $\Lambda_A$ as follows
\be
i \bar{\psi} \dsl \psi  &\longrightarrow& 
i \bar{\psi} \dsl \psi - i \vec{\Theta} \left( \bar{\psi} \, i \partial_\mu
\gamma^\mu  \gmf  
\frac{\vec{\tau}}{2} \, \psi +  \bar{\psi}  \, \gmf \frac{\vec{\tau}}{2} i 
 \partial_\mu \gamma^\mu \, \psi \right)
\\
& = &   i \bar{\psi} \dsl \psi
\ee
The second term vanishes because $\gmf$ anti-commutes with $\gamma_\mu$.
Thus the Lagrangian is also invariant under $\Lambda_A$ with the conserved 
`axial - vector' current 
\be
A_{\mu }^a = \bar{\psi} \gamma_\mu \gmf \frac{\tau}{2} \psi
\label{ax_current}
\ee
In summary, the Lagrangian of massless fermions, and, hence, massless QCD, 
is invariant
under  both transformations, $\Lambda_V$ and $\Lambda_A$.\footnote{Note, that 
the above  
Lagrangian is also invariant under the operations $\psi \rightarrow 
exp(-i \Theta) \psi$ and  $\psi \rightarrow exp(-i \gmf \Theta) \psi$. The
first operation is related to the conservation of the baryon number while the
second symmetry is broken on the quantum level. This is referred to as the U(1)
axial anomaly, which is a real breaking of the symmetry in contrast to
the spontaneous breaking discussed below (see
e.g. \cite{DGH92}).} 
This symmetry
is what is meant by chiral symmetry\footnote{Often, people talk about `chiral'
symmetry but actually only refer to the axial transformation
$\Lambda_A$. This is due to its special role is plays, since it is 
 spontaneously broken in the ground state.}.
The chiral symmetry is often referred to by its group structure as the 
$SU(2)_V \times SU(2)_A$ symmetry.

Now let us see, what happens if we introduce a mass term.
\be
\delta {\cal L} = - m \, (\bar{\psi} \psi)
\ee
From the above, $\delta {\cal L}$ is obviously invariant under the vector
transformations $\Lambda_V$ but {\em not} under $\Lambda_A$
\be
\Lambda_A: \,\, m \, (\bar{\psi} \psi) \longrightarrow m \bar{\psi} \psi 
-2 i m \vec{\Theta} \left( \bar{\psi} \frac{\vec{\tau}}{2} \gmf \psi \right)   
\ee
Thus, $\Lambda_A$ is not a good symmetry, if the fermions (quarks) have a 
finite
mass. But as long as the masses are small compared
to the relevant scale of the theory one may treat $\Lambda_A$ as an approximate
symmetry, in the sense, that predictions based under the assumption of the
symmetry should be reasonably close to the actual 
results\footnote{A wheel which
is slightly bent and thus not perfectly invariant under rotations, can for 
most practical purposes still be considered as being round, as long as the
bending is small compared to the radius of the wheel.}.

In case of QCD we know that the masses of the light quarks are about $5-10 \,
\rm MeV$ whereas the relevant energy scale given by $\Lambda_{QCD} \simeq 200 
\, \rm MeV$ is considerably larger. We, therefore, expect that $\Lambda_A$ 
should be an approximate symmetry and that the axial current should be 
approximately (partially) conserved. This slight symmetry breaking due to the
quark masses is the basis of the so called Partial Conserved Axial Current 
hypothesis (PCAC). Furthermore, as long as the symmetry breaking is small, one 
would also expect, that its effect can be described in a perturbative approach.
This is carried out in a systematic fashion in the framework of chiral
perturbation theory.

\subsection{Chiral Symmetry and PCAC}

\subsubsection{Chiral transformation of mesons}
In order to develop a better feeling for the meaning of the symmetry
transformations $\Lambda_V$ and $\Lambda_A$, let us find out how pions and
rho-mesons transform under these operations. To this end, let us consider 
combinations of quark fields, which carry the quantum numbers of the mesons
under consideration. This should give us the correct transformation
properties:\\

\noindent
\begin{center}
pion-like state: $\vec{\pi} \equiv i \bar{\psi} \vec{\tau} \gmf \psi$;
\hspace{1cm}  
sigma-like state: \ \ \ $\sigma \equiv \bar{\psi} \psi$ \ \ \ \ \ \ \\\
rho-like state: \ $\vec{\rho}_\mu \equiv \bar{\psi} \vec{\tau} \gamma_\mu 
\psi$; \hspace{1cm}
$a_1$-like state: \ \ \ \ $\vec{a_1}_\mu \equiv \bar{\psi} \vec{\tau} 
\gamma_\mu \gmf \psi$\\
\end{center}
Here, the vector again indicates 
the iso-vector nature of the mesons such as pion 
and rho etc. i.e. these particles transform like a vector under 
{\em isospin} - rotations.  In addition particles, which transform like 
vectors under a Lorentz transformation, have an additional Lorentz index 
$\mu$. These are the vector  mesons  $\rho$ and $a_1$, 
which carry a total  spin of one.

\ \\
\noindent
{\bf (i)} vector transformations $\Lambda_V$, see 
eqs. (\ref{vec_1},\ref{vec_2}): 
\be
\pi_i: \,\,\,  i \bar{\psi} \tau_i \gmf \psi &\longrightarrow& 
i \bar{\psi} \tau_i \gmf \psi  +  \Theta_j \left(\bar{\psi} \tau_i \gmf
\frac{\tau_j}{2} \psi -  \bar{\psi} \frac{\tau_j}{2} \tau_i \gmf \psi \right)
\nonumber \\ 
& = &   i \bar{\psi} \tau_i \gmf \psi + i \Theta_j \epsilon_{ijk} \, 
\bar{\psi} \gmf \tau_k \psi
\ee
where we have used the commutation relation between the $\tau$ matrices
$[\tau_i,\tau_j] = 2 i \epsilon_{ijk} \tau_k$.
In terms of pions this can be written as
\be
\vec{\pi} \longrightarrow \vec{\pi} + \vec{\Theta} \times \vec{\pi}
\label{vec_pi}
\ee
which is nothing else but an isospin rotation, namely the isospin direction of
the pion is rotated by $\Theta$. The same result one obtains for
the $\rho$ - meson
\be
\vec{\rho_\mu} \longrightarrow \vec{\rho_\mu} + \vec{\Theta} \times 
\vec{\rho_\mu}
\ee
Consequently, the vector-transformation $\Lambda_V$ can be identified with 
the isospin
rotations and the conserved vector current with the isospin current, which we
know to be conserved in strong interactions.\\

\noindent
{\bf (i)} axial transformations $\Lambda_A$, see eqs. (\ref{ax_1},\ref{ax_2}): 
\be
\pi_i: \,\,\,  i \bar{\psi} \tau_i \gmf \psi &\longrightarrow& 
i \bar{\psi} \tau_i \gmf \psi + \Theta_j \left(\bar{\psi} \tau_i \gmf \gmf
\frac{\tau_j}{2} \psi  +  \bar{\psi} \gmf \frac{\tau_j}{2} \tau_i \gmf \psi 
\right)
\nonumber \\ 
& = &   i \bar{\psi} \tau_i \gmf \psi + \Theta_i \bar{\psi} \psi
\ee
where we have made use of the anti-commutation relation of the $\tau$ matrices
$\{\tau_i,\tau_j \} = 2 \delta_{ij}$ and  of $\gmf \gmf = 1$.
In terms of the mesons this reads:
\be
\vec{\pi} \longrightarrow \vec{\pi} + \vec{\Theta} \sigma
\label{ax_pi}
\ee
and similarly for the $\sigma$-meson
\be
\sigma \longrightarrow \sigma - \vec{\Theta} \vec{\pi}
\label{ax_sigma}
\ee
The pion and the sigma-meson  are obviously rotated into each other  under 
the axial transformations 
$\Lambda_A$. Similarly the rho rotates into the $a_1$
\be
\vec{\rho}_\mu \longrightarrow \vec{\rho}_\mu + 
\vec{\Theta} \times \vec{a_1}_\mu
\label{ax_rho}
\ee
Above we just have convinced 
ourselves that $\Lambda_A$ is a symmetry of the QCD
Hamiltonian. Naively, this would imply, that states which can be rotated into
each other by this symmetry operation should have the same Eigenvalues, i.e the
same masses. This, however, is clearly not the case, since $m_\rho = 770 \, \rm
MeV$ and  $m_{a_1} = 1260 \, \rm MeV$. We certainly do not expect that the 
slight
symmetry breaking due to the finite current quark masses is responsible for
this splitting. This should lead to mass differences which are small
compared to the masses themselves. In case of the $\rho$ and $a_1$, however,
the mass difference is of the same order as the mass of the $\rho$.  
The resolution to this problem will be the spontaneous breakdown of the axial
symmetry. Before we discuss what is meant by that, let us first convince 
ourselves, that the axial vector is conserved to a good approximation, so that
the axial symmetry must be present somehow.

\subsubsection{Pion decay and PCAC}
Let us first consider the weak decay of the pion. In the simple Fermi
theory the weak interaction Hamiltonian  is of the current-current
type, where both currents are a sum of axial and vector currents, as we have
defined them above (see e.g. \cite{CB83}). 
Because of parity, the weak
decay of the pion is controlled by the matrix element of the axial 
current between the
vacuum and the pion, $<0 | A_\mu | \pi>$. This matrix element must be
proportional to the pion momentum, because this is the only vector around
\be
<0 | A_\mu^a(x) | \pi^b(q) > = -i f_\pi q_\mu \delta^{ab} e^{-iq \cdot x}
\label{pcac1}
\ee
and the proportionality constant $f_\pi = 93 MeV$ is determined from
experiment\footnote{The are several definitions of $f_\pi$ around, depending on
whether factors of 2, $\sqrt{2}$ are present in eq. \raf{pcac1}.}. Here, 
the indices a and b refer to isospin whereas $\mu$ again indicates the 
Lorentz vector character of the axial current.
 
Let us now
take the divergence of eq. \raf{pcac1} 
\be
<0 | \partial^\mu A_\mu^a(x) | \pi^b(q) > = 
- f_\pi q^2 \delta^{ab} e^{-iq \cdot x} = - f_\pi m_\pi^2 \delta^{ab} 
e^{-iq \cdot x}
\label{pcac2}
\ee
To the extent, that the pion mass is small compared to hadronic scales,
the axial current is approximately conserved. Or in other words, the smallness
of the pion mass is directly related to the partial conservation of the axial
current, i.e. to the fact that the axial transformation is an approximate
symmetry of QCD. In the literature the above relation \raf{pcac2} is often 
referred to as the PCAC relation. The above relations 
(\ref{pcac1},\ref{pcac2}) also
suggest, that the axial current carried by a pion is
\be
A_{\mu,\, pion}^a  = f_\pi \partial_\mu \Phi^a(x)
\label{pcac3}
\ee
or that the divergence of the axial-vector current can be identified with the
pion field (up to a constant). Here $\Phi^a(x)$ is the pion field. 
Sometimes this relation between pion field
and axial current is also referred to as the PCAC relation.\\

\subsubsection{Goldberger-Treiman relation}
There is more evidence for the conservation of the axial current. Let us
consider the axial current of a nucleon. This is simply given by (see eq. 
\raf{ax_current})
\be
A_{\mu,\, nucleon}^a = g_a \bar{\psi_N} \, \gamma_\mu \gmf \frac{\tau^a}{2} 
\, \psi_N
\label{a_n}
\ee
where $\psi_N = (proton,neutron)$ is now an iso-spinor representing proton and
neutron. The factor $g_a = 1.25$, is due to the fact, that the axial current 
of the  nucleon is renormalized by $25 \%$, as seen in the weak beta decay of
the neutron.
Since the nucleon has a large mass $M_N$, we do
not expect that its axial current is conserved, and indeed by using the free
Dirac equation for the nulceon one can show that
\be
\partial^\mu A_{\mu,\, nucleon}^a = 
i g_a  M_N \, \bar{\psi_N} \gmf \tau^a \psi_N \neq 0
\label{a_n2}
\ee
which vanishes only in case of vanishing nucleon mass. 
We know, however, that the nucleon interacts strongly with the pion. Therefore,
let us assume that the total axial current is the sum of the nucleon and the
pion contribution. Using the PCAC-relation \raf{pcac3} and equ. \raf{a_n}
we have 
\be
A_\mu^a = g_a \bar{\psi_N} \gamma_\mu \gmf \frac{\tau^a}{2} \psi_N +
f_\pi \partial_\mu \Phi^a
\ee
If we require, that the total current is conserved, 
$\partial^\mu A_\mu = 0$, we obtain 
\be
\partial^\mu \partial_\mu \Phi^A = - g_a \, i \frac{M_N}{f_\pi} 
\bar{\psi_N} \gmf \tau^a \psi_N
\ee
where we have used \raf{a_n2}. This is nothing else but a Klein Gordon
equation for a massless boson (pion) coupled to the nucleon. Hence, requiring 
the
conservation of the total axial current immediately leads us to the prediction 
that
the pion should be massless. This is exactly what we also concluded from the
weak pion decay. If we now allow for a finite pion mass, which is equivalent to
requiring that the divergence of the axial current is consistent with the PCAC
result \raf{pcac2}, then we arrive at the Klein Gordon equation for a pion
coupled to the nucleon
\be
\left( \partial^\mu \partial_\mu  + m_\pi^2 \right) \Phi = 
- g_a \, i \frac{M_N}{f_\pi} \bar{\psi_N} \gmf \tau \psi_N
\ee
where the pion-nucleon coupling constant is given by
\be
g_{\pi NN} = g_a \frac{M_N}{f_\pi} \simeq 12.6
\label{gt}
\ee
This is to be compared with the value for the pion-nucleon coupling as
extracted e.g. from pion-nucleon scattering experiments
\be
g_{\pi NN}^{exp} = 13.4
\ee
which is in remarkably close agreement, considering the fact, that equ. 
\raf{gt}
relates the strong-interaction pion-nucleon coupling $g_{\pi NN}$ with
quantities extracted from the weak interaction, namely $g_a$ and $f_\pi$. Of
course, the reason why this works is that there is some symmetry, namely chiral
symmetry,  at play, which
allows to connect seemingly different pieces of physics. 
Equation  \raf{gt} is usually called the Goldberger-Treiman relation.

\subsubsection{Spontaneous breakdown of chiral symmetry}
There appears to be some contradiction: On the one hand the meson mass 
sepctrum does not reflect the axial-vector symmetry. On the other hand,
the weak pion decay seems to be consistent with a (partially) conserved  
axial-vector current. Also the success of the Goldberger-Treiman relation
indicates that the axial-vector current is conserved and, hence, that the axial
transformation $\Lambda_A$ is a symmetry of the strong interactions.

The solution to this puzzle is, that the axial-vector symmetry is {\em 
spontaneously} broken. What does one mean by that? One speaks of a  
spontaneously
broken symmetry, if a symmetry of the Hamiltonian is not realized in the
ground state.

\begin{figure}
\setlength{\epsfysize}{7in}
\centerline{\epsffile{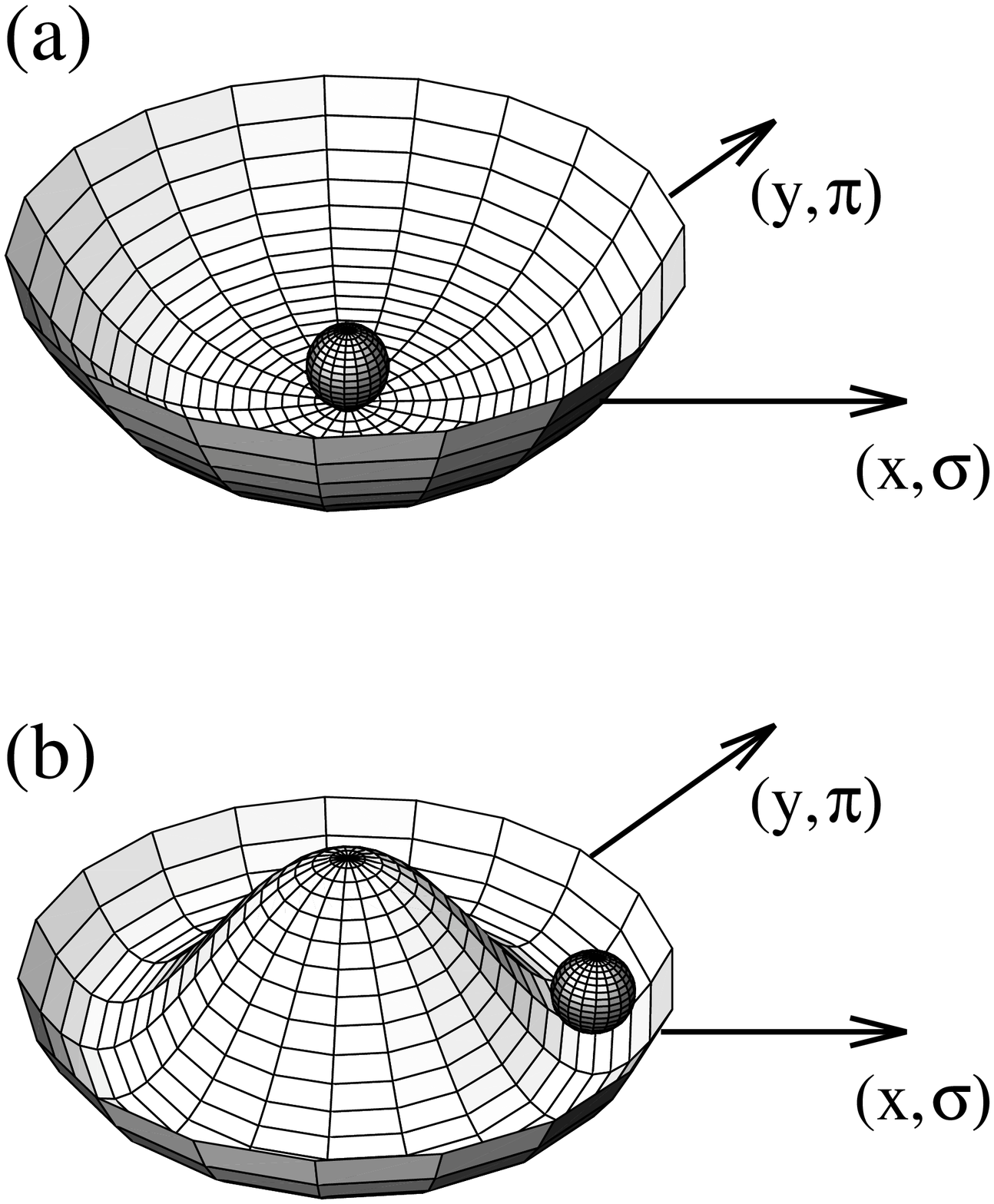}}
\caption{Effective potentials. (a) No spontaneous breaking of symmetry. (b)
Spontaneous breaking of symmetry.}
\label{sponti}
\end{figure}

This is best illustrated in a classical
mechanics analog. In fig. \ref{sponti} we have two rotationally invariant
potentials (`interactions'). 
In (a) the ground state is right in the middle, and
the potential plus ground state are still invariant under rotations. In (b), on
the other hand, the ground state is at a finite distance away from the center.
The point at the center is a local maximum of the potential and thus unstable.
If we put a little ball in the middle, it will roll down somewhere and 
find its groundstate some place in the valley which represents the
true minimum of the potential. By picking one point in this valley (i.e picking
the ground state), the rotational symmetry is obviously broken. Potential plus
groundstate are not symmetric anymore. The symmetry has been broken
{\em spontaneously} by choosing a certain direction to be the groundstate. 
However, 
effects of the symmetry are still present. Moving the ball around in the
valley (rotational excitations) does not  cost any energy, whereas radial
excitations do cost energy.
 
Let us now use this mechanics analogy in order to understand what the
spontaneous breakdown of the axial-vector symmetry of the strong interaction
means. Assume, that the effective QCD-hamiltonian  at zero temperature has 
a form similar to that depicted
in fig. \ref{sponti}(b), where the (x,y)-coordinates are replaced by
$(\sigma,\vec{\pi})$-fields. The spacial rotations are then the mechanics  
analog of the 
axial-vector rotation $\Lambda_A$, which rotates $\vec{\pi}$ into $\sigma$ (see
equ. \raf{ax_pi}). Since the ground state is not at the center but
a some finite distance away from it, one of the fields will have a finite
expectation value. This can only be the  $\sigma$-field, because it carries the
quantum numbers of the vacuum. In the quark language, this means
we expect to  have a finite scalar quark condensate $< \bar{q}q > \neq 0$.
In this picture, 
pionic excitation correspond to small 'rotations' away from the 
ground-state along the valley, which do not cost any energy. 
Consequently the mass of the pion should be zero.
In other words, due to the spontaneous breakdown of
chiral symmetry, we predict a vanishing pion mass.
Excitations in the
$\sigma$-direction correspond to radial excitations and therefore are
massive. 

This scenario is in perfect agreement with what we have found above. 
The spontaneous breakdown of the axial-vector symmetry leads to different
masses of the pion and sigma. However, since the interaction itself is
still symmetric, pions  become massless, which is exactly 
what we find from the PCAC relation, provided that the axial current is
perfectly conserved. 
Thus the mesonic mass spectrum as well as the PCAC--
and the Goldberger-Treiman relation are consistent with a spontaneous breakdown
of the axial-vector symmetry $\Lambda_A$. The pion appears as a massless mode
(Goldstone boson) as a result of the symmetry of the interaction.

Incidentally, the assumption of a spontaneously broken axial-vector symmetry also explains the mass difference between the $\rho$- and
$a_1$ meson and one predicts that $m_{a_1} =  \sqrt{2} m_\rho$ in good
agreement with the measured masses. The derivation of this
result, however, is too involved to be presented 
here and the interested reader is referred to the literature 
\cite{Wei67,Sak69}.

One expects, that at high temperature/densities the finite expectation value of
the scalar quark condensate melts away resulting in  a system, where chiral
symmetry is not spontaneously broken anymore. In this, as it is often called, 
chirally restored phase pion/sigma as well as rho/$a_1$, if they
exist\footnote{If deconfinement and chiral restoration occur at the same
temperature, it may become meaningless to talk about mesons above the critical temperature.}, 
should be degenerate
and the pion looses its identity as a Goldstone boson, i.e. it will become
massive. The effective interaction in this phase would then have a shape
similar to fig \ref{sponti}(a). It is one of the major goals of the
ultrarelativistic heavy ion program to create and identify a macroscopic 
sample of this phase in the laboratory.

In the following section we will construct a  chiral invariant 
Lagrangian, the so called `Linear-sigma-model', 
in order to see how the
concept of spontaneous breakdown of chiral symmetry is realized in the
framework of a simple model. We will also
discuss  how to incoorporate the effect of the finite quark masses leading to 
the explicit breaking of chiral symmetry.

\section{Linear sigma-model}
\subsection{Chiral limit}
In this section we will construct a simple chirally invariant model involving
pions and nucleons, the so
called linear sigma - model. 
This model was first introduced by Gell-Mann and Levy in 1960 \cite{GL60}, 
long before QCD was known to be the theory of the strong interaction.
In order to construct such a model, we have to
write down a Lagrangian which is a Lorentz-scalar and which is invariant
under the vector- and axial-vector transformations, $\Lambda_V$ and
$\Lambda_A$. 

In the previous section, we have shown, that
the pion transforms under  $\Lambda_V$ and
$\Lambda_A$  as (\ref{vec_pi},\ref{ax_pi}).
\be
\Lambda_V: \pi_i \longrightarrow \pi_i + \epsilon_{ijk} \Theta_j \pi_k
\hspace{1cm}
\Lambda_A: \pi_i \longrightarrow \pi_i + \Theta_i \sigma
\label{2.1}
\ee
Similarly one can also show, that the $\sigma$-field transforms like
\be
\Lambda_V: \sigma \longrightarrow \sigma
\hspace{1cm}
\Lambda_A: \sigma \longrightarrow \sigma - \Theta_i \pi_i
\label{2.2}
\ee
Since $\Lambda_V$ is simply an isospin rotation, the squares of the fields
are  invariant under this transformation
\be
\Lambda_V: \;\; \pi^2 \longrightarrow \pi^2; \
\hspace{1cm} 
\sigma^2 \longrightarrow \sigma^2
\ee
whereas under $\Lambda_A$ they transform like
\be
\Lambda_A: \;\;  \vec{\pi}^2 \longrightarrow \vec{\pi}^2 + 
2 \sigma \Theta_i \pi_i ; \
\hspace{1cm} 
\sigma^2 \longrightarrow \sigma^2 - 2 \sigma \Theta_i \pi_i
\ee
However, the combination $(\vec{\pi}^2 + \sigma^2)$ is invariant under both
transformations, $\Lambda_V$ {\em and} $\Lambda_A$
\be
(\vec{\pi}^2 + \sigma^2) \stackrel{\Lambda_V, \Lambda_A} \longrightarrow 
(\vec{\pi}^2 + \sigma^2)
\label{inv}
\ee
Since this combination is also a Lorentz-scalar, we can build a chirally
invariant Lagrangian around this structure:
\begin{itemize}
\item Pion-nucleon interaction:\\
The standard pion nucleon interaction involves a pseudo-scalar combination of
the nucleon field multiplied by the pion field:
\be
g_\pi \left( i \bar{\psi} \gmf \vec{\tau} \psi \right) \, \vec{\pi} 
\ee
where from now on we denote the pion-nucleon coupling constant simply 
by $g_\pi$.
Under the chiral transformations this transforms exactly like $\pi^2$, because
the term involving the nucleon has the same quantum numbers as the pion.
Chiral invariance requires that there must be another term, which
transforms like $\sigma^2$, in order to have the  invariant structure
\raf{inv}. The simplest choice is a term of the form,
\be
g_\pi \left( \bar{\psi} \psi \right) \, \sigma 
\label{n_sigma}
\ee
so that the interaction term between nucleons and the mesons is
\be
\delta {\cal L} = - g_\pi \left[  (i \bar{\psi} \gmf \vec{\tau} \psi) \, 
\vec{\pi}   + (\bar{\psi} \psi) \, \sigma \right]
\label{pn_int}
\ee
\item Nucleon mass term:\\
We know that an explicit nucleon mass term breaks chiral invariance (see
section \ref{massless} ). The nucleon mass is also too large to be simply a
result of the small explicit chiral symmetry breaking as reflected in the PCAC
relation \raf{pcac2}.
The simplest\footnote{Actually one can allow for an
explicit nucleon mass term if one also includes the chiral partner of the
nucleon, which is believed to be the $N^*(1535)$. This is an interesting
alternative approach which is discussed in detail in ref. \cite{dTK89}} way to
give the nucleon a mass without breaking chiral symmetry, is to exploit the
coupling of the nucleon to the $\sigma$-field \raf{n_sigma}, which has the
structure of a nucleon mass term. This, however, requires that the 
$\sigma$-field as a {\em finite
vacuum expectation value},
\be
<\sigma> = \sigma_0 = f_\pi
\label{sig_vac}
\ee
where the choice of $\sigma_0 = f_\pi$ is dictated by the Goldberger-Treiman
relation \raf{gt} in the limit of $g_a = 1$. A finite vacuum expectation
value for the $\sigma$-field immediately implies, that chiral symmetry will
be spontaneously broken, as discussed in the last section. In order for our
model to generate such an expectation value, we have to introduce a potential
for the sigma field, which has its minimum at $\sigma  = f_\pi$. This brings
us to the next ingredient of our model.
\item Pion - sigma potential:\\
The potential, which generates the vacuum expectation value of the $\sigma$
field has to be a function of the invariant structure \raf{inv} in order to be
chirally invariant. The simplest choice is:
\be
V = V(\pi^2 + \sigma^2) = \frac{\lambda}{4} \left( (\pi^2 + \sigma^2) - 
f_\pi^2 \right)^2
\label{pot}
\ee
This potential, which is plotted in fig. \raf{pot_fig} (see also
fig. (\ref{sponti}b) for a three-dimensional view ) indeed has its minimum at
$\sigma = f_\pi$ for $\pi = 0$. Due to its shape, it is often referred to as
the `Mexican - hat - potential'.
\item Kinetic energy terms:\\
Finally we have to add kinetic energy terms for the nucleons and the mesons
which have the form $i \bar{\psi} \dsl \psi$ and $\frac{1}{2} ( \partial_\mu
\pi \partial^\mu \pi + \partial_\mu \sigma \partial^\mu \sigma ) $,
respectively. 
Both are
chirally invariant. The first term is just the Lagrangian of free mass less
fermions, which we have shown to be invariant. The second term again has the
invariant structure \raf{inv}.
\end{itemize}

\begin{figure}[t]
\setlength{\epsfysize}{4in}
\centerline{\epsffile{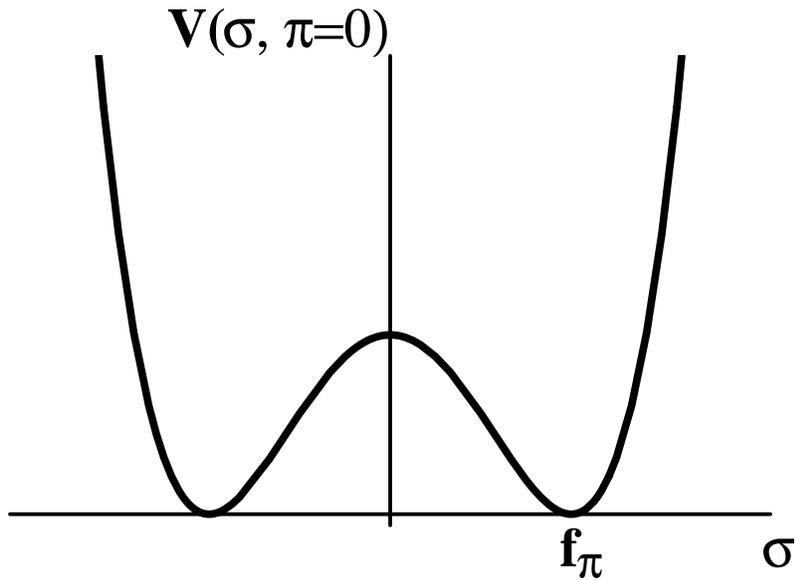}}
\caption{Potential of linear sigma-model}
\label{pot_fig}
\end{figure}

Putting everything together, the 
Lagrangian of the linear sigma-model reads (remember that the
potential V enters with a minus-sign into the Lagrangian):
\be
{\cal L}_{L.S.} & = & i \bar{\psi} \dsl \, \psi  - 
g_\pi \left(  i \bar{\psi} \gmf \vec{\tau} \psi \, \vec{\pi}
+ \bar{\psi} \psi \, \sigma \right)
\nonumber \\
&  &
- \frac{\lambda}{4} \left( (\pi^2 + \sigma^2) - f_\pi^2 \right)^2
+ \frac{1}{2} \partial_\mu \pi \partial^\mu \pi + 
\frac{1}{2}\partial_\mu \sigma \partial^\mu \sigma 
\label{L_LS}
\ee

What are the properties of this model? Let us start with the ground state. 
As already mentioned, in the ground state the $\sigma$ - field has a finite
expectation value, whereas the pion has none, because of parity. Furthermore,
the nucleon obtains its mass  from its  interaction  with
the sigma field. But what are the masses of the $\sigma$ and $\pi$ - mesons?
There are no explicit mass terms for the $\sigma$- and $\pi$-fields in the 
Lagrangian \raf{L_LS}, but, as with the
nucleon, there could be some coupling to the expectation value of the $\sigma$
field, which gives rise to mass terms. From the structure of the potential (see
figs \raf{pot_fig} and \raf{sponti}) as well as from our discussion of the
spontaneous breakdown of chiral symmetry, we expect the pion to be massless
and the $\sigma$-meson to become massive. In order to verify that, 
let us expand
the potential \raf{pot} for small fluctuations  around the ground state.
\be
\sigma = \sigma_0 + \dsig; \,\,\,\, \pi = \dpi
\label{expand}
\ee
Actually, it is these  fluctuations ($\dsig, \, \dpi$), which  are to 
be be identified with the observed
particles ($\sigma$- and $\pi$- meson). 
Since a bosonic mass term is quadratic in the fields (see Lagrangian 
\raf{L_KG}), let us
expand the potential up to quadratic order in the fluctuations $\dsig \, \dpi$.
Expanding around a minimum, the linear order vanishes, and we have:
\be
V(\sigma,\pi) = \lambda f_\pi^2 \dsig^2 + {\cal O}(\delta^3)
\ee
where we have used that $\sigma_0 = f_\pi$. Comparing with the Lagrangian of a
free boson we identify the mass of the sigma to be (remember that $L = T-V$)
\be
m_\sigma^2  =  2 \lambda f_\pi^2 \neq 0
\label{m_sig}
\ee
We find no mass term for the pion in agreement with our expectation,
that the pion should be the massless Goldstone boson of the 
spontaneously broken
chiral symmetry. 

In summary, the properties of the ground state of the linear
sigma-model are:

\be
< \sigma > &=& \sigma_0 = f_\pi\\
<\pi> &=& 0 \\
M_N &=& g_\pi \sigma_0 = g_\pi f_\pi \\
m_\sigma^2 &=&  2 \lambda f_\pi^2 \neq 0 \\
m_\pi  &=& 0
\label{ground}
\ee

Before we conclude this section, let us calculate the conserved axial current
and check, if the PCAC-relation is satisfied in our model.
The infinitesimal axial transformations of the nucleon, pion and sigma fields
are given by (see \raf{ax_1}, \raf{2.1} and \raf{2.2})
\be
\psi &\longrightarrow& \psi - i \gmf \frac{\tau^a}{2} \Theta^a \psi
\\
\pi^i &\longrightarrow& \pi^i + \Theta^a \delta^{i,a} \sigma
\\
\sigma &\longrightarrow& \sigma - \Theta^a \pi^a
\ee
Comparing with the general form \raf{trans} for unitary
transformations, we find that the generator of the axial transformation $T^a$ 
act on the fields in the following way
\be
T^a \psi & = & \gmf \frac{\tau^a}{2}  \psi
\\
T^a \pi^j & = & i \sigma \delta^{a,j}
\\
T^a \sigma & = & - i \pi^a 
\ee
Using the expression for the conserved current \raf{conserved} the
conserved axial current is given by
\be
A_\mu^a =  \bar{\psi} \gamma_\mu \gamma_5 \frac{\tau^a}{2} \psi 
- \pi^a \partial_\mu
\sigma + \sigma \partial_\mu \pi^a
\label{axial_sig}
\ee

In order to check the PCAC-relation, we again expand the fields around
the ground state (see eq. \raf{expand})
\be
A_\mu^a = \bar{\psi} \gamma_\mu \gamma_5 \frac{\tau^a}{2} \psi - 
(\delta \pi^a) \partial_\mu \dsig
 + \dsig \partial_\mu (\delta \pi^a) + f_\pi \partial_\mu (\delta \pi^a)
\label{axial_sig2}
\ee
where we have used that $\sigma_0 = f_\pi$. Since the PCAC-relation involves
the 
matrix element $<0 | A_\mu^a | \pi^j>$ only the last term of \raf{axial_sig2}
contributes. The other terms would require either nucleons or 
sigma-mesons in the final or  initial state. Thus, as far as the PCAC relation
is concerned, the axial current reduces to $(\dpi = \pi)$ 
\be
A_\mu^a(x)_{PCAC} = f_\pi \partial_\mu \pi(x)
\ee
in agreement with the PCAC-results eq. \raf{pcac3}.

\subsection{Explicit breaking of chiral symmetry}
\label{sec_3.2}
So far we have assumed that the axial-vector symmetry is a perfect symmetry of
the strong interactions. From our discussion in section
\ref{massless} we know, however, 
that the small but finite current quark masses of the up and
down quark break the axial-vector symmetry explicitly. This {\em explicit} 
breaking
of the symmetry should not be confused with the {\em spontaneous} 
breakdown, we have
discussed before. In case of a spontaneous breaking of a symmetry the
Hamiltonian is still symmetric, whereas in case of an explicit breaking,
already the Hamiltonian is not  symmetric.

One may wonder if the whole concept of spontaneous symmetry breaking makes
any sense if already the Hamiltonian is not symmetric. The answer to that,
again, depends on the scales involved. If the explicit symmetry breaking is 
small, i.e. if the quark masses are small compared to to relevant energy scale
of QCD, as we believe they are, then it 
will be sensible to apply the notion of a
spontaneously broken symmetry. 

To illustrate that, let us again utilize our little mechanics analogy, which 
we have developed in the previous section. An explicit symmetry breaking
would imply that both potentials of figure \raf{sponti} are not 
invariant under rotation. This could for instance be achieved by
slightly tilting them towards, say, the x-direction. As a result, also 
the  ground state of potential (a) is away from the center ($x,y = 0$). But
the dislocation is small compared to that due to the spontaneous
breaking. Furthermore, as long as the potentials are tilted only slightly,
rotational excitation (pions) in potential (b) are still
considerably softer than the radial ones (sigma-mesons).  So in this sense, we
expect the effect due to the spontaneous breakdown of chiral symmetry to
dominate the dynamics, as long as the explicit breaking is small. 
In the linear sigma-model, the mass scale generated by the spontaneous
breakdown is the nucleon mass, whereas that generated by the explicit
breakdown will be the mass of the pion, as we shall see.  Thus, indeed the
explicit breaking is small, and our picture, developed under the assumption of
perfect axial-vector symmetry, will survive the introduction of the explicit
breaking to a very good approximation.

After these remarks let us now introduce a symmetry breaking term into the
linear sigma-model. In QCD, we know, that the symmetry is explicitly broken
by a quark mass-term
\be
\delta{\cal L}_{X \chi SB} = - m \bar{q}q
\label{sb_qcd}
\ee
where the subscript $X\chi SB$ stands for explicit chiral symmetry breaking.
If we identify, as we have done before, the scalar quark-field combination
$\bar{q}q$ with the $\sigma$ field, this would suggest the following
symmetry breaking term in the sigma-model
\be
\delta{\cal L}_{SB} = \epsilon \sigma
\label{sb}
\ee
where $\epsilon$ is the symmetry breaking parameter. 
This term clearly is not invariant under the axial transformation $\Lambda_A$
but preserves the vector symmetry $\Lambda_V$.
Including this
term, the potential $V$ \raf{pot} now has the form
\be
V(\sigma, \pi) = \frac{\lambda}{4} \left( (\pi^2 + \sigma^2) - v_0^2 \right)^2
  - \epsilon \sigma
\label{pot_sb}
\ee
where we now have replaced $f_\pi$ of eq. \raf{pot} by a general 
parameter $v_0$, which in limit of $\epsilon \rightarrow 0$ will go to $f_\pi$.
The effect of the symmetry breaking term is to tilt the potential slightly
towards the positive $\sigma$ direction, and thus to break the symmetry (see
fig. \raf{pot_fig_sb}).

\begin{figure}[t]
\setlength{\epsfysize}{4in}
\centerline{\epsffile{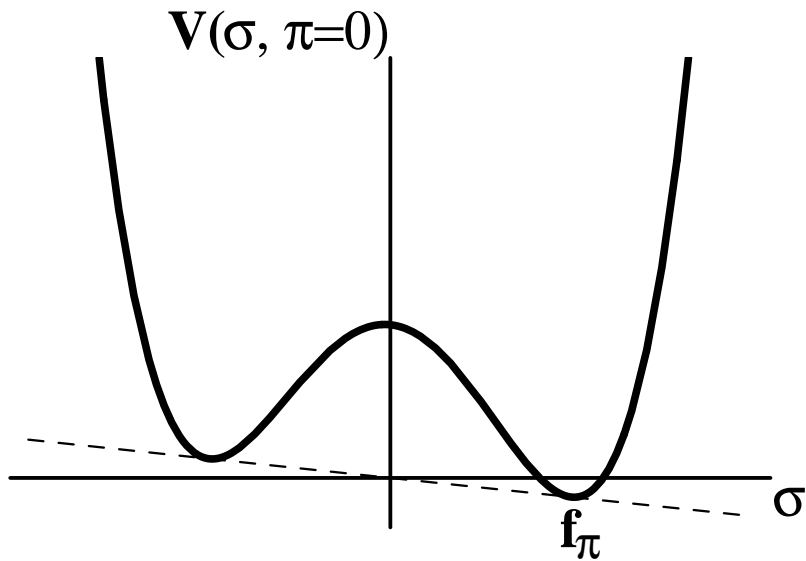}}
\caption{Potential of linear sigma-model with explicit symmetry breaking}
\label{pot_fig_sb}
\end{figure}

What are the consequences of this additional term? First of all, the minimum
has shifted slightly. If we require that the value of the new minimum is still
$f_\pi$ in order to preserve the Goldberger-Treiman relation, we find for the
parameter $v_0$ to leading order in $\epsilon$
\be
v_0  = f_\pi - \frac{\epsilon}{2 \lambda f_\pi^2}
\label{eq:v_0}
\ee
Also the mass of the sigma is slightly changed
\be
m_\sigma^2 = \left. \frac{\partial^2 V}{\partial \sigma^2} \right|_{\sigma_0} =
2 \lambda f_\pi^2 + \frac{\epsilon}{f_\pi}
\label{msig_sb}
\ee
But most importantly, the pion now acquires a finite mass
\be
m_\pi^2 = \left. \frac{\partial^2 V}{\partial \pi^2} \right|_{\sigma_0} =
\frac{\epsilon}{f_\pi} \neq 0
\label{mpi}
\ee
which fixes the parameter $\epsilon$
\be
\epsilon = f_\pi m_\pi^2
\ee
Thus, the square of the pion mass  is directly proportional to the symmetry 
breaking parameter $\epsilon$ as we would have expected it from our previous
discussion. 

Due to our choice of $\sigma_0 = f_\pi$, the nucleon mass is not changed,
which, however, does not mean that there is 
no contribution to the nucleon mass from
the explicit symmetry breaking.  If we split the nucleon mass into a
contribution from the symmetric part of the potential ($\sim v_0$) and one from
the symmetry breaking term ($\sim \epsilon$),
\be
M_N = g_\pi \sigma_0 = g_\pi \, (v_0 + \frac{\epsilon}{2 \lambda f_\pi^2})
\ee
we find that 
the contribution from the symmetry breaking, which is often referred to as the
pion-nucleon sigma-term\footnote{This definition of the pion-nucleon sigma term
should be taken with some care. For a rigorous definition see e.g. 
\cite{Hoe83,AFF73}. In the framework of the
sigma-model, this definition, however, is correct to leading order in
$\epsilon$.}, is given by
\be
\Sigma_{\pi N} = \delta M_N^{X \chi SB} = g_\pi \frac{\epsilon}{2 \lambda
f_\pi^2} \simeq g_\pi f_\pi \frac{m_\pi^2}{m_\sigma^2}
\label{sig_pin}
\ee
As we shall see below, the pion-nucleon sigma-term can be measured in
pion-nucleon scattering experiments and its is currently believed to be
\cite{Hoe83} 
$\Sigma_{\pi N}(0) = 35 \, \pm 5 \rm MeV $.

Since chiral symmetry is now explicitly broken, the axial-vector current is
not conserved anymore. 
The functional form of the axial current is
the same, however,  as in the symmetric case, eq. \raf{axial_sig},
because the symmetry breaking term \raf{sb} does not involve any derivatives 
(see equ. \raf{conserved}). Its divergence is 
related to the variation of the symmetry breaking term in the Lagrangian,
as shown at the end of section \ref{symmetries}.
\be
\partial^\mu A_\mu^a =  \epsilon \, \delta (\sigma) = - f_\pi m_\pi^2 \pi^a
\ee
which leads directly to the PCAC relation \raf{pcac2}. Here $\delta (\sigma)$
denotes the variation of the $\sigma$-field with respect to the axial-vector
transformation $\Lambda_A$, not the fluctuation around the ground state. 
As in equ. \raf{conserved} the angel $\Theta^a$ has
been divided out.

The main effect of the explicit chiral symmetry breaking was to give the pion a
mass. But we can utilize the symmetry breaking further to derive\footnote{These
`derivations' are merely heuristic, but I feel they nicely demonstrate  
the physics which
is going on. For a rigorous derivation see e.g. \cite{AFF73}.} some rather
useful relations between expectation values of the scalar quark operator 
$\bar{q}q$ and measurable quantities like $f_\pi$, $m_\pi$, and $\Sigma_{\pi
N}$. 

When we introduced the symmetry breaking term into our model, we had
required that it has the same transformation properties under the chiral
transformations as the QCD-symmetry breaking term. The overall strength of the
symmetry breaking, $\epsilon$ we then adjusted to reproduce the ground state
properties, namely  the pion mass. Therefore, it seems reasonable to expect,
that that the vacuum expectation value of the
symmetry breaking terms in QCD \raf{sb_qcd} and in the effective model \raf{sb}
are the same. 
\be
<0| \, \epsilon \sigma \, |0> & = & <0| - m \bar{q} q |0>
\ee
If we insert for $\epsilon = m_\pi^2 f_\pi$ and use $<0| \sigma |0> = f_\pi$ 
we arrive at the so called
Gell-Mann -- Oakes -- Renner (GOR) relation \cite{AFF73,GOR68}
\be
m_\pi^2 f_\pi^2 = - \frac{m_u + m_d}{2} <0| \bar{u}u + \bar{d} d|0>
\label{GOR}
\ee
where we have written out explicitly 
the average quark mass, $m$, and the quark operator $\bar{q}q$. The GOR 
relation is extremely useful, since it relates the quark condensate with 
$f_\pi$ and/or the pion mass with the current-quark mass. 

Similarly, but less convincingly, one can argue, that the contribution to the 
nucleon mass due to chiral symmetry breaking, $\Sigma_{\pi N}$, is the
expectation value of the symmetry breaking Hamiltonian 
$\delta H_{X\chi SB} = - \delta {\cal L}_{X\chi SB}$ 
between nucleon states. This leads to the exact expression of the
pion-nucleon sigma-term in terms of QCD variables \cite{Hoe83,GL82}
\be
\Sigma_{\pi N} = \frac{m_u + m_d}{2} <N| \bar{u}u + \bar{d} d|N>
\label{sig_pin2}
\ee
This relation will turn out to be very helpful in order to estimate the change
of the chiral condensate in nuclear matter at finite density.

\subsection{S-wave pion-nucleon scattering}
\label{pn_scatter}
In order to see how chiral symmetry affects 
the dynamics, let us, as an example,
study pion-nucleon scattering in the sigma-model. Let us begin by
introducing some notation. 
\begin{figure}[thb]
\setlength{\epsfysize}{1.5in}
\centerline{\epsffile{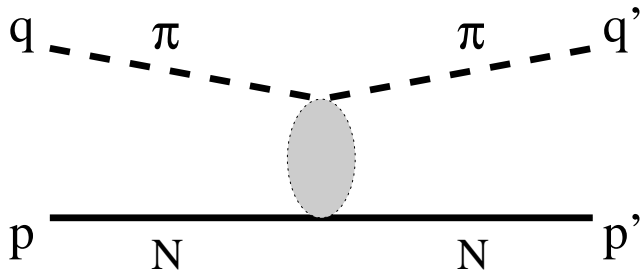}}
\caption{Pion nucleon scattering amplitude.}
\label{diagram1}
\end{figure}

The invariant
scattering amplitude $T(q,q')$ is commonly decomposed into
a scalar and a vector part\footnote{For details see
e.g. the appendix of \cite{EW88}.}
 (see fig. \raf{diagram1} for the notation of
momenta) 
\be
T(q,q') = A(s,t) + \frac12 \gamma^\mu (q_\mu + q_\mu') B(s,t)
\label{amp0}
\ee
where $(s,t)$ are the usual Mandelstam variables, and $q$ and $q'$ denote the
incoming and outgoing pion - four-momenta. 
The relativistic scattering amplitude is related to the more familiar
scattering amplitude in the center of mass frame, ${\cal F}(\vec{q},\vec{q'})$
by
\be
\chi^+ {\cal F} \chi' = \frac{M_N}{4 \pi \sqrt{s}} \bar{u}(p,\sigma) T 
u(p',\sigma')
\label{amp1}
\ee
Here $\chi$ are Pauli-spinors for the nucleon representing spin and isospin and
$u(p,\sigma)$ 
stands for a relativistic spinor of a nucleon with momentum $p$ and
spin $\sigma$

The scattering amplitude can be decomposed into isospin-even and -odd
components\footnote{Notice, that the isospin-odd amplitude is the {\em
negative} of what in the literature is commonly called the iso-vector 
amplitude whereas the isospin-even amplitude is identical to the so called
isoscalar one (see \cite{EW88}).}
\be
T_{ab} = T^+ \delta_{ab} + \frac12 [\tau_a,\tau_b] T^-
\label{amp2}
\ee
where the indices $a,b$ refer to the isospin.

In the discussion of pion-nucleon scattering instead of (s,t) one usually 
uses the invariant variables \cite{Hoe83}
\be
\nu & = &\frac{s-u}{4 M_N}\\
\nu_B & = & - \frac{1}{2 M_N} q^\mu q_\mu' = \frac{1}{4 M_N} (t - q^2 - q'^2)
\label{amp3}
\ee
The spin-averaged, non-spin-flip $(\sigma = \sigma')$, 
forward ($p = p'$) scattering 
amplitude, which will be most relevant for
the aspects of chiral symmetry, is usually denoted by $D$ and is given in terms
of the above variables by
\be
D \equiv \frac12 \sum_\sigma \bar{u}(p,\sigma) T u(p,\sigma) = A  + \nu B
\label{amp4}
\ee
Finally, if one wants to extract effects due to explicit chiral symmetry
breaking, one best analyses the so called subtracted amplitude
\be
\bar{D} = D - D_{PV} = D - \frac{g_\pi^2}{M_N} \frac{\nu_B^2}{\nu_B^2 - \nu^2}
\label{dbar}
\ee

Now let us calculate the pion-nucleon scattering amplitude in the sigma-
model. At tree level the diagrams shown in fig. \raf{diagram2} contribute
to the amplitude. The first two processes represent the 
simple absorption and re-emission of the pion by the nucleon. Provided,
that there is a coupling between pion and nucleon, one would have written down
these diagrams immediately, without any knowledge of chiral symmetry. 
The third diagram (c), which involves the exchange of a sigma-meson,  is a 
direct result of chiral symmetry, and, as well shall see, is crucial in order 
to give the correct value for  the amplitude.

\begin{figure}[thb]
\setlength{\epsfxsize}{5.5in}
\centerline{\epsffile{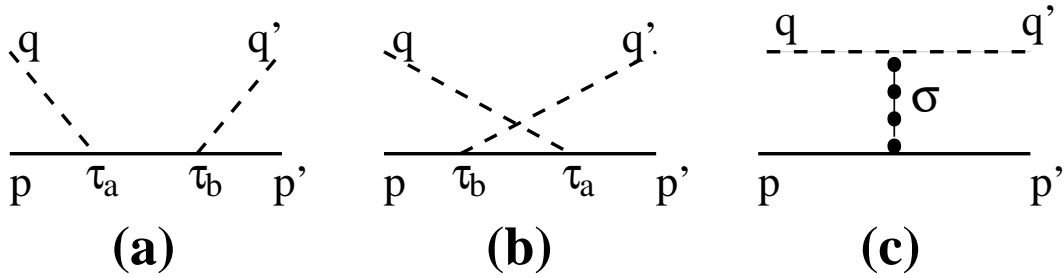}}
\caption{Diagrams contributing to the pion-nucleon scattering amplitude
$T_{ab}$.} 
\label{diagram2}
\end{figure}

In the following, we will restrict ourselves to the forward scattering
amplitudes, i.e. $q = q'$ and $p = p'$.
Using standard Feynman-rules (see e.g. \cite{BD65}), the above diagrams
can be evaluated in a straightforward fashion. For diagram (a) we obtain  
\be
\lefteqn{\bar{u}(p) T_{ab}^{(a)} u(p) } 
\nonumber \\ 
& = & g_\pi^2 \bar{u}(p) \, \tau_a \gmf 
\frac{ (p+q)^\mu \gamma_\mu + m}{(p+q)^2 - m^2} \tau_b \gmf \, u(p)
\nonumber \\ 
& = & \bar{u}(p) \left[ (\delta_{ab} + \frac12 [\tau_a,\tau_b] ) 
(- g_\pi^2 \frac{ q^\mu \gamma_\mu}{s - m^2} ) \right] \,u(p)
\ee
where we have used that $\gmf \gamma_\mu = - \gamma_\mu \gmf$, $\gmf^2 = 1$,
$\tau_a \tau_b = \delta_{ab} +\frac12 [\tau_a,\tau_b]$, and the Dirac equation
$(p_\mu \gamma^\mu - m) u(p) = 0$.
Obviously, diagram (a) contributes only to the vector piece of the amplitude,
$B$, and the isospin-even and -odd amplitudes are the same
\be 
B_{(a)}^+ = B_{(a)}^- = - \frac{g_\pi^2}{s - M_N^2} 
\ee
The contribution of the crossed or u-channel (diagram (b)) one 
obtains by replacing
\be
s &\rightarrow& u
\\
(\tau_a \tau_b) &\rightarrow& (\tau_b \tau_a)
\\
q &\rightarrow& - q
\ee
with the result
\be
B_{(b)}^+ = \frac{g_\pi^2}{u - M_N^2} = -B_{(b)}^-
\ee
Here isospin-even and -odd amplitudes have the opposite sign.  

It is instructive to calculate the scattering amplitude resulting from the
first two diagrams only. If we didn't know about chiral symmetry, and, hence, 
the
existence of the $\sigma$-exchange diagram, this is what we would naively
obtain. At threshold $(\vec{q} = 0)$, the combined amplitudes are
\be
\nu B_{(a)+(b)}^+ & = &- \frac{g_\pi^2}{M_N} 
\left( \frac{1}{1 - \frac{m_\pi^2}{4 M_N^2}} \right)
\label{b_ab_pl}
\\
\nu B_{(a)+(b)}^- & = & 
g_\pi^2 \frac{m_\pi}{2 M_N^2} \left( \frac{1}{1 - \frac{m_\pi^2}{4 M_N^2}}
\right)
\ee
Using equations (\ref{amp0}, \ref{amp1}) the resulting 
s-wave isospin-even and isospin-odd scattering length, $a_0$, 
which is related to the scattering amplitude D \raf{amp4} 
at threshold by
\be
a^\pm_0 = \frac{1}{4 \pi (1 + \frac{m_\pi}{M_N})} D^\pm_{\rm at \; threshold}
\ee
would be 
\be
a_0^+((a)+(b)) 
& = & - \frac{g_\pi}{4 \pi f_\pi (1 + \frac{m_\pi}{M_N})} \, 
(1 + {\cal O}(\frac{m_\pi^2}{M_N^2}) )  
\simeq - 1.4 \, m_\pi^{-1}
\\
a_0^-((a)+(b)) 
& = &  \frac{m_\pi}{8 \pi f_\pi^2 (1 + \frac{m_\pi}{M_N})} \, 
(1 +  {\cal O}(\frac{m_\pi^2}{M_N^2}) ) 
\simeq  0.078 \,m_\pi^{-1}
\label{a_odd}
\ee
where we have made  of the Goldberger-Treiman relation $g_\pi f_\pi = M_N$.
This is to be compared with the experimental values of \cite{EW88}
\be
a_0^+ (exp) = -0.010(3) \, m_\pi^{-1} \hspace{1cm} 
a_0^- (exp) =  0.091(2) \, m_\pi^{-1}
\ee

While we find reasonable  agreement for the isospin-odd amplitude, the isospin
even amplitude is off by two orders of magnitude! A different
choice of the pion-nucleon coupling $g_\pi$ would not fix the problem, but just
shift it from one amplitude to the other.
Before we evaluate the remaining diagram (c), let us point out that in the
chiral limit, i.e. $m_\pi= 0 $, the isospin-odd amplitude vanishes.

In order to evaluate the $\sigma$-exchange diagram, we need to extract the
pion-sigma coupling from our Lagrangian. This is done by expanding the
potential $V$ \raf{pot} up to third order in the field fluctuations ($\dpi$ and
$\dsig$). The terms proportional to $\sim \dpi^2 \dsig$ then give the desired
coupling. 
\be
\delta {\cal L}_{\pi \pi \sigma} = - \lambda f_\pi \dpi^2 \dsig
\ee
The resulting amplitude is then given by
\be
\bar{u}(p) T_{ab}^{(c)} u(p) = - g_\pi \frac{ 2 \lambda f_\pi}{t - m_\sigma^2}
\delta_{ab}
\label{sig_exchange}
\ee
It only contributes to the scalar part of the amplitude, A, and only in the
isospin-even channel.
Using  $2 \lambda f_\pi^2 = m_\sigma^2 - m_\pi^2$ (see
eqs. (\ref{msig_sb}, \ref{mpi}) ) we find
\be
A^+_{(c)} = - \frac{g_\pi}{f_\pi} \frac{m_\sigma^2 - m_\pi^2}{m_\sigma^2-t}
=  \frac{g_\pi}{f_\pi} \left( 1  - \frac{t - m_\pi^2}{t-m_\sigma^2} \right)
\ee
To leading order, the contribution to the s-wave scattering lengths 
of diagram (c) is
\be
a_0^+((c)) 
& = & \frac{g_\pi}{4 \pi f_\pi (1 + \frac{m_\pi}{M_N})} 
(1 + {\cal O}(\frac{m_\pi^2}{m_\sigma^2}) )  
\\
a_0^-((c)) & = & 0 
\ee
Thus, to leading order, the contribution of the $\sigma$-exchange diagram (c) 
{\em exactly} cancels that of the nucleon-pole diagrams ((a) and (b))
and the total isospin-even scattering length vanishes
\be
a_0^+ = 0 + {\cal O}(\frac{m_\pi^2}{M_N^2},\frac{m_\pi^2}{m_\sigma^2})
\ee
in much better agreement with experiment. 
The cancelation between the large
individual contributions to the isospin-even amplitude is a direct 
consequence of chiral symmetry, which required the $\sigma$-exchange diagram. 
In the chiral limit, this cancelation is
perfect, i.e. the isospin-even scattering amplitude vanishes identically,
because the corrections $\sim m_\pi$ are zero in this case.

Furthermore, since the third diagram (c) does not contribute to the isospin-odd
amplitude, the good agreement found above still holds. 
In other words, with the `help' of chiral symmetry both amplitudes are
reproduced well.

Putting all terms together  the isospin-even
amplitude $D^+$ is given in terms of the variables $\nu$ and $\nu_B$
\be
D^+(\nu,\nu_B) & = & A^+ + \nu B^+
\nonumber \\ 
& = & \frac{g_\pi}{f_\pi} \frac{\nu^2}{\nu_B^2 - \nu^2}
  + \frac{g_\pi}{f_\pi} \left( 1  -
  \frac{t-m_\pi^2}{t-m_\sigma^2} \right)
\nonumber \\ 
& = &       \frac{g_\pi}{f_\pi} \frac{\nu_B^2}{\nu_B^2 - \nu^2}
 - \frac{g_\pi}{f_\pi}\frac{t-m_\pi^2}{t-m_\sigma^2} 
\ee
Here the first term in the second 
line is the contribution form diagrams (a) and
(b) and the other two term are from diagram (c). 
At threshold, where $\nu = m_\pi$, $\nu_B = -\frac{m_\pi^2}{2 M_N}$, and 
$t = 0$ this reduces to 
\be
D^+_{\rm at \; threshold} = - \frac{g_\pi}{f_\pi} \left( 
\frac{m_\pi^2}{4 M_N^2 - m_\pi^2} + \frac{m_\pi^2}{m_\sigma^2} \right)
\label{totamp}
\ee
As already pointed out, to leading order ($\sim m_\pi^0$) or in 
the chiral limit, this amplitude vanishes, as a result chiral symmetry. However
the contribution next to leading order $\sim m_\pi^2$ involve also the mass of
the $\sigma$-meson, which has not yet been clearly identified
in experiment. In the sigma-model, this mass essentially is a free
parameter, since it is directly proportional to the coupling $\lambda$. Since 
$\lambda$ gives the strength of the invariant potential $V$, chiral 
symmetry considerations will not determine this parameter.
Thus, aside from the very important finding, that the isospin-even scattering
length should be small, the linear sigma-model as no predictive power for
the {\em actual} small value of the scattering length\footnote{In the framework
of chiral perturbation theory, the value of the isospin-even amplitude is
essentially regarded as an input to fix the parameters of the expansion. There
are attempts to relate the value of the scattering length to
contributions from the Delta \cite{BKM93}. In this approach, the problem
is shifted to the determination of an  unknown off-shell parameter  appearing
in the Delta-propagator.}. 

Notice, that although D is the spin averaged, forward $(t = 0)$ scattering 
amplitude,  we can obviously study it an any value of $\nu$, 
$t$ or equivalently
$\nu$ and $\nu_B$. A kinematical point of particular interest is the so called
Cheng-Dashen point, given by
\be
\nu = 0, \hspace{1cm} t = 2 m_\pi \rightarrow \nu_B = 0
\ee
At this kinematical point , the subtracted amplitude $\bar{D}$ \raf{dbar}
is directly related with the pion-nucleon sigma-term $\Sigma_{\pi N}$
\cite{Hoe83} 
\be
\bar{D}(\nu = 0, t = 2 m_\pi)=  \frac{\Sigma_{\pi N}}{f_\pi^2} 
\ee
In the sigma-model we find for the subtracted amplitude to leading order in the
pion mass
\be
\bar{D}(\nu = 0, t = 2 m_\pi) = -\frac{g_\pi}{f_\pi}
\frac{m_\pi^2}{m_\sigma^2} = \frac{\Sigma_{\pi N}}{f_\pi^2} 
\ee
where we have used the expression for the sigma-term, derived
above \raf{sig_pin} from the contribution of the explicit chiral symmetry
breaking to the nucleon mass.
Notice, although the Cheng-Dashen point is in an unphysical region, it can
nevertheless be reached via dispersion relation techniques, and, thus, the
sigma-term can be extracted from pion-nucleon scattering data. For a detailed
discussion, see ref. \cite{Hoe83}.

\section{Nonlinear sigma-model}
One of the disturbing features of the linear sigma-model is the existence of
the $\sigma$-field, because it cannot really be identified with any existing
particle. Furthermore, at low energies and temperatures one would expect that
excitations in the $\sigma$-direction should be much smaller than pionic
ones, which in the chiral limit are massless (see fig. \raf{sponti}).
This is supported by our results for the pion-nucleon scattering, where in the
final result the mass of the sigma-meson only showed up in next to leading
order corrections, which vanish in the chiral limit. 

Let us, therefore, remove  the $\sigma$-meson as a dynamical field by sending
its mass to infinity. Formally this can be achieved by assuming an infinitely
large coupling $\lambda$ in the linear sigma-model. As a consequence the
mexican-hat potential gets infinitely steep in the sigma-direction (see
figure below ). This   confines
the dynamics to the circle, defined by the  minimum of the potential.
\be
\sigma^2 + \pi^2 = f_\pi^2
\label{constraint}
\ee 

\begin{figure}[h]
\setlength{\epsfxsize}{5.5in}
\centerline{\epsffile{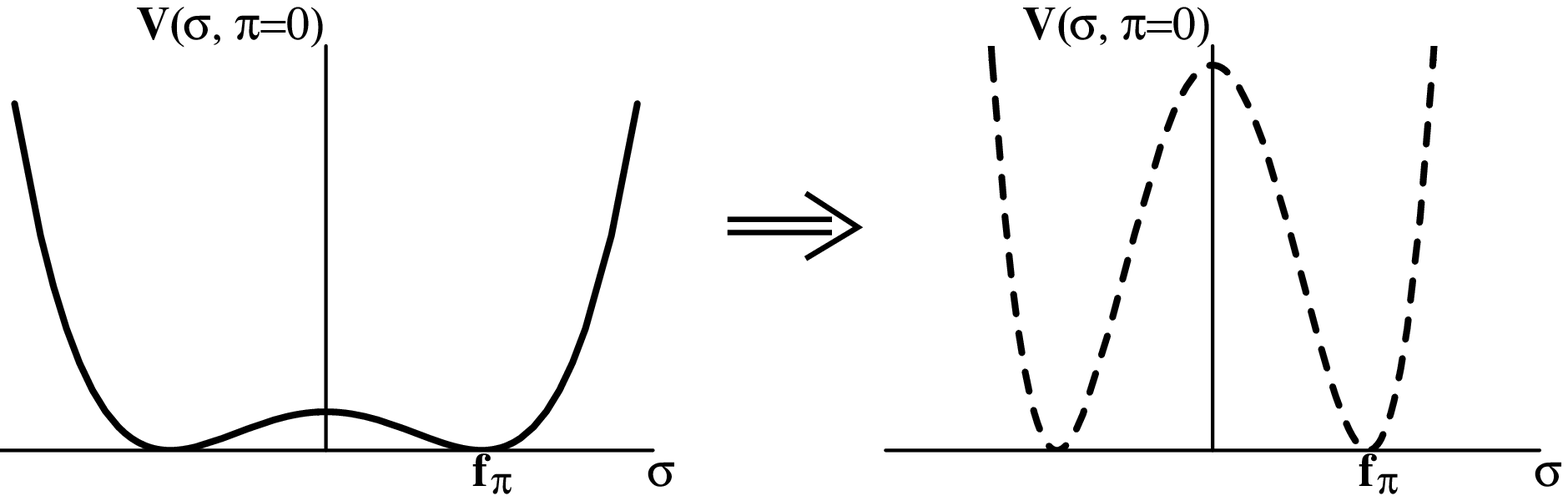}}
\label{steep}
\end{figure}

This additional condition removes one degree of freedom, which close to the
ground state, where $<\sigma> = f_\pi$, is  the sigma field, and we are
left with pionic excitations only. Because of the above constraint
\raf{constraint}, the dynamics is now restricted to rotation on the so called
chiral circle (actually it is a sphere). Therefore, 
the fields can be expressed in terms of angles $\vec{\Phi}$,
\be
\sigma(x) & = & f_\pi \cos(\frac{\Phi(x)}{f_\pi}) = f_\pi + {\cal O}( \Phi^2)
\nonumber \\ 
\vec{\pi} (x) & = & 
f_\pi \hat{\Phi} \sin(\frac{\Phi(x)}{f_\pi}) = \vec{\Phi}(x) +
{\cal O}( \Phi^3)
\ee
which to leading order can be identified with the pion field. 
Here, $\Phi = \sqrt{\vec{\Phi}\vec{\Phi}}$ and $\hat{\Phi} =
\frac{\vec{\Phi}}{\Phi}$. Clearly, this
ansatz fulfills the  constraint \raf{constraint}. Equivalently, one can chose
a complex notation for the fields, as it is commonly done in the literature
\be
U(x) = e^{i \frac{\vec{\tau} \vec{\Phi}(x)}{f_\pi}} = 
\cos(\frac{\Phi(x)}{f_\pi}) + 
i \vec{\tau} \hat{\Phi} \sin(\frac{\Phi(x)}{f_\pi}) =
\frac{1}{f_\pi} (\sigma + i \vec{\tau} \vec{\pi})
\ee
where $U$ represents a unitary $(2 \times 2)$ matrix. The constraint
\raf{constraint} is then equivalent to
\be
\frac{1}{2} tr (U^+ U) = \frac{1}{f_\pi^2} (\sigma^2 + \pi^2) = 1
\label{uu}
\ee
Since chiral symmetry corresponds to
a  symmetry with respect to rotation around the chiral circle, 
all structures of the form
\be
tr (U^+U), \; tr (\partial_\mu U^+ \partial^\mu U) \ldots
\ee
are invariant. 
Already at this point it 
becomes obvious that we eventually will need some scheme,
which tells us which structures to include and which ones not. This will lead 
us to the ideas of chiral perturbation theory in the following section.

Let us continue by rewriting the Lagrangian of the linear sigma-model
\raf{L_LS} in terms of the new variables $U$ or $\Phi$. After a little algebra
we find that the kinetic energy terms of the mesons is given by
\be
\frac{1}{2}\partial_\mu \sigma \partial^\mu \sigma +
\frac{1}{2}\partial_\mu \vec{\pi} \partial^\mu \vec{\pi} 
= \frac{f_\pi}{4} tr(\partial_\mu U^+ \partial^\mu U)
\ee
Next, we realize that nucleon-meson coupling term can be written as  
\be
- g_\pi \left(  \bar{\psi} \psi \, \sigma  
+ \bar{\psi} \gmf \vec{\tau} \psi \, \vec{\pi} \right)
& = & -g_\pi  \bar{\psi} \left[ f_\pi \left( \cos(\frac{\Phi}{f_\pi}) + 
i \gmf \vec{\tau} \hat{\Phi} \sin(\frac{\Phi}{f_\pi}) \right) \right] \psi
\nonumber \\ 
& = & - g_\pi \bar{\psi} \left(f_\pi e^{i \gmf \frac{\vec{\tau} 
\vec{\Phi}(x)}{f_\pi}} \right)  \psi  
\nonumber \\ 
& = & - g_\pi f_\pi \bar{\psi} \Lambda \Lambda  \psi
\label{inter}
\ee
where we have defined
\be
\Lambda \equiv e^{i \gmf \frac{\vec{\tau} \vec{\Phi}(x)}{2 f_\pi}}
\ee
If we now redefine the nucleon fields 
\be
\psi_W  & = & \Lambda \psi
\\
\Rightarrow \bar{\psi}_W & = & \psi^+ \Lambda^+ \gamma^0  
\stackrel{\{\gamma_0, \gmf \}=0}{ = }    
\psi^+ \gamma^0  \Lambda = \bar{\psi} \Lambda
\ee
the interaction term \raf{inter} can be simply written as
\be
-g_\pi f_\pi \bar{\psi} \Lambda \Lambda  \psi = 
-g_\pi f_\pi \bar{\psi}_W \psi_W  =   - M_N \bar{\psi}_W \psi_W
\ee
where we have used the Goldberger-Treiman relation \raf{gt}. In terms of the
new fields, $\psi_W$, 
the entire  interaction term as been reduced to the nucleon
mass term. If we want to identify the nucleons with the redefined fields 
$\psi_W$ we also have to rewrite the nucleon kinetic energy term in terms of
those fields.
\be
\bar{\psi}\, i \dsl \, \psi = \bar{\psi}_W \Lambda^+ i \dsl \, \Lambda^+ \psi_W
\ee
Since $\Lambda$ is space-dependent through the fields $\Phi(x)$, the derivative
also acts on $\Lambda$, giving rise to additional terms.
After some straightforward algebra, one finds 
\be
\bar{\psi}_W \Lambda^+ i \dsl \, \Lambda^+ \psi_W = 
\bar{\psi}_W \left( i \dsl + \gamma^\mu V_\mu + \gamma^\mu \gmf A_\mu \right)  
\psi_W
\ee
with
\be
V_\mu & = &\frac12 \left[ \xi^+ \partial_\mu \xi + \xi \partial_\mu \xi^+
\right]
\label{vector}
\\
A_\mu & = &\frac{i}{2} \left[ \xi^+ \partial_\mu \xi - \xi \partial_\mu \xi^+
\right]
\label{axial}
\\
\xi & = & e^{i \frac{\vec{\tau} \vec{\Phi}(x)}{2 f_\pi}} \Rightarrow U 
= \xi \xi
\ee

We do not need to transform the potential of the linear
sigma-model, $V(\pi,\sigma)$, since it vanishes on the chiral circle
due to the constraint condition \raf{constraint}. Putting everything together,
the Lagrangian of the nonlinear sigma-model, which is often referred to as
the Weinberg-Lagrangian, reads in the above variables
\be
{\cal L}_W = \bar{\psi} \left( i \dsl + \gamma^\mu V_\mu + 
\gamma^\mu \gmf A_\mu - M_N \right) \psi +
\frac{f_\pi}{4} tr(\partial_\mu U^+ \partial^\mu U) 
\label{wein_1}
\ee
were we have dropped the subscript from the nucleon fields.
Clearly, this Lagrangian  depends nonlinearly on the fields $\vec{\Phi}$.
It is instructive to expand the Lagrangian for small fluctuations $\Phi /
f_\pi \ll 1$ around the ground state. This gives
\be
{\cal L}_W & \simeq & \bar{\psi}( i \dsl - M_N) \psi + 
\frac12 ( \partial_\mu \vec{\Phi})^2 
\nonumber \\ 
&& + \frac{1}{2 f_\pi} (\bar{\psi} 
\gamma_\mu \gmf \vec{\tau} \psi) \partial^\mu
\vec{\Phi} 
-  \frac{1}{4 f_\pi^2} (\bar{\psi} \gamma_\mu \vec{\tau} \psi) \cdot 
\left( \vec{\Phi} \times (\partial^\mu \vec{\Phi}) \right) 
\label{wein_2}
\ee
where $\vec{\Phi}$ is now to be identified with the pion field.
Comparing with the linear sigma-model, the $\sigma$-field has disappeared
and the coupling between nucleons and pions has been changed to a
pseudo-vector-one, involving the derivatives (momenta) of the
pion-field. In addition, an explicit isovector coupling-term has emerged.
From this Lagrangian it is immediately clear that the s-wave pion-nucleon
scattering amplitudes vanishes in the chiral limit, because all couplings 
involve the pion four-momentum, which at threshold is zero in case of massless
pions. Thus, the important cancelation between the nucleon pole-diagrams and 
the $\sigma$-exchange diagram, which we found in the linear sigma-model, 
has been moved into the derivative coupling of the pion through the above
transformations.   

On the level of the expanded Lagrangian \raf{wein_2}, the explicit breaking of
chiral symmetry is introduced by an explicit pion mass term.
Consequently corrections to the scattering lengths due to the nucleon pole
diagrams should be of the order of $m_\pi^2$, since two derivative couplings
are involved.  However, the coupling  
$ \delta {\cal L} = -  \frac{1}{4 f_\pi^2} 
(\bar{\psi} \gamma_\mu \vec{\tau} \psi ) \cdot \left( \vec{\Phi} \times
(\partial^\mu \vec{\Phi}) \right) $, 
which 
contributes to first order to the isospin-odd amplitude, should give rise to
a term $\sim \frac{m_\pi}{f_\pi^2}$ in agreement with our previous
findings \raf{a_odd}. Not too surprisingly one finds, that the above Lagrangian
gives exactly the same results for the scattering-length as the linear 
sigma-model, except, that corrections $\sim \frac{1}{m_\sigma^2}$ are absent, 
because
we have assumed that the mass of the $\sigma$-meson is infinite. However, the
full Lagrangian \raf{wein_1} would give rise to many more terms, if we expand
to higher orders in the fields $\Phi$, which then would lead to loops etc.
How to control these corrections in a systematic fashion 
will be the subject of the
following section, where we discuss the ideas of chiral perturbation theory.

\section{Basic ideas of Chiral Perturbation Theory}

In the previous sections we were concerned with the most simple chiral
Lagrangian in order to see how chiral symmetry enters into the dynamics. As we
have already pointed out, many more chirally invariant terms can be 
included into the Lagrangian and thus we need some scheme which tells us 
what to include and what not. This scheme is provided by chiral perturbation 
theory. 

Roughly speaking, the essential idea of chiral perturbation theory is to
realize that at low energies the dynamics should be controlled by the lightest
particles, the pions, and the symmetries of QCD, chiral symmetry. 
Therefore, s-matrix elements, i.e. scattering amplitudes,
should be expandable in a Taylor-series of the pion-momenta and masses, 
which is also consistent with chiral symmetry.  This scheme will
be valid until one encounters a resonance, such as the $\rho$-meson, which
corresponds to a singularity of the s-matrix. Practically speaking, above the 
resonance, a Breit-Wigner distribution cannot be expanded in a Taylor series.

It is not too surprising that such a scheme works. Imagine, we did not know
anything about QED. We still could go ahead and parameterize the, say, 
electron-proton scattering amplitude in powers of the momentum transfer $t$. 
In this
case the Taylor coefficients would be related to the total charge, the charge
radius etc.  With this information we could write down an effective
proton-electron Lagrangian, where the  
couplings are fixed by the above Taylor-coefficients, namely the charge and the
charge- radius. This effective theory will, of course, reproduce the
results of QED up to the order, which has been fixed by experiment. 
It is in this sense, the effective Lagrangian, obtained in
chiral perturbation theory, should be understood; namely as a method
of writing s-matrix elements to a given order in pion-momentum/mass. 
And to the order considered, the the effective Lagrangian
obtained with chiral-perturbation theory should be equivalent with QCD 
\cite{Wei79,Leu95a}.

It should be stressed, that chiral perturbation theory is not a
perturbation theory in the usual sense, i.e.,  it is not a 
perturbation theory in the QCD-coupling constant. 
In this respect, it is actually a
nonperturbative method, since it takes already infinitely many orders of the 
QCD
coupling constant in order to generate a pion. Instead, 
as already pointed out, 
chiral perturbation theory is an expansion of the s-matrix elements 
in terms of pion-momenta/masses.

From the above arguments one could get the impression, that chiral perturbation
theory has no predictive power, since it represents simply  a power expansion
of measured scattering amplitudes. Although this may true in some cases, 
one could
easily imagine that one fixes the effective Lagrangian from some experiments
and then is able to calculate other observables. For example, 
imagine that the effective pion-nucleon interaction has been fixed 
from pion nucleon-scattering experiments. This interaction can then be used to
calculate e.g. the photo-production of pions.

To be specific, let us discuss the case of pure pionic interaction,
i.e. without any nucleons. 
As pointed out in the previous section, chiral invariance requires 
that
the effective Lagrangian has to be build from structures involving $U^+U$
\raf{uu} such as
\be
tr (\partial_\mu U^+ \partial^\mu U), \;\; tr (\partial_\mu U^+ \partial^\mu U)
tr (\partial_\mu U^+ \partial^\mu U), \;\; 
tr[ (\partial_\mu U^+ \partial^\mu U)^2], \, \ldots
\ee
Furthermore, each $U = e^{i \frac{\vec{\tau} \vec{\Phi}(x)}{f_\pi}}$ contains
any power of the pion-field $\Phi$, which may 
give rise to loops etc. To specify,
which of the above terms should be included into the effective Lagrangian and
how much each term should be expanded in terms of the pion field, one has to
count the powers of pion momenta contributing to the desired process
(scattering amplitude).

Consider a given Feynman-diagram contributing to the scattering amplitude. It
will have a certain number $L$ of loops,  a certain number $V_i$ of vertices of
type $i$ 
involving $d_i$ derivatives of the pion field an a certain number of internal
lines $I_p$. The power $D$ of the pion
momentum $q$, this diagram will have at the end, can be determined as follows:
\begin{itemize}
\item each loop involves an integral over the internal momenta $\int d^4 q \sim
q^4$
\item each internal pion line corresponds to a pion propagator, and thus
contributes as $\frac{1}{q^2}$
\item each vertex $V_i$ involving $d_i$ derivatives of the pion field, 
contributes
like $q^{d_i}$
\end{itemize}
Consequently, the total power of $q$, $q^D$ is given by
\be
D = 4 L - 2 I_P + \sum_i V_i d_i
\ee
This can be simplified by using the general relation between the numbers of
loops, internal lines and vertices of a given diagram
\be
L = I_p - \sum_i V_i + 1
\ee
to give
\be
D = 2 + 2 L + \sum_i V_i (d_i - 2)
\label{count_pi}
\ee
With this formula we can determine to which order of the Taylor expansion of
the scattering amplitude a given diagram contributes. 

In order to see how this counting rule leads to an effective Lagrangian of a
given order, we best study the simple example of pion-pion scattering.
Since $U^+ U = 1$ does not contribute to the dynamics, the simplest
contribution to the effective Lagrangian is given by
\be
{\cal L}_{2} = \frac{f_\pi}{4} tr(\partial_\mu U^+ \partial^\mu U)
\label{d_2}
\ee
where the subscript denotes the number of derivatives involved. 
Since we are discussing pion-pion scattering, we have to expand at least up to 
fourth order in the pion fields,
\be
{\cal L}_{2} = \frac12 (\partial_\mu \Phi)^2 + 
\frac{1}{6 f_\pi^2} \left[ (\Phi \partial_\mu \Phi)^2 - 
\Phi^2 (\partial_\mu \Phi \partial^\mu \Phi) \right] + {\cal O}(\Phi^6)
\ee 
where the second term contributes to the pion-pion scattering amplitude.
Although this term has two contributions, 
for the purposes of power counting, the second term may be considered as 
one vertex function, because both contributions have the same 
number of derivatives. Thus, to lowest order, we have just one diagram, which 
is shown in fig. \raf{f_pipi_0}. It has no loops, $L=0$, and the vertex
function carries two derivatives of the pion field. Using the above counting
rule \raf{count_pi}, the order of this diagram is $D=4$. 

We can easily convince
ourselves that there are no more terms contributing to this order. Including
terms into the Lagrangian with four derivatives of the pions field such as e.g.
$ tr[ (\partial_\mu U^+ \partial^\mu U)^2]$ immediately leads to $D \geq 6$. 
Also expanding the above Lagrangian \raf{d_2} up to sixth order in the pion
field leads to $D\geq 6$, because two of the pion fields have to be combined
into a loop, since we are only considering  a process
with four external pions. 

Obviously, the order of the effective Lagrangian depends on the process
under consideration. Whereas a term involving six pion fields contributes to
the order $D \geq 6$ to pion-pion scattering, it would contribute to order
$D=4$ to a process with three initial and three final pions. Of course, having
realized, that we are actually parameterizing s-matrix elements, this is not 
such a surprise.

\begin{figure}[t]
\setlength{\epsfxsize}{1.5in}
\centerline{\epsffile{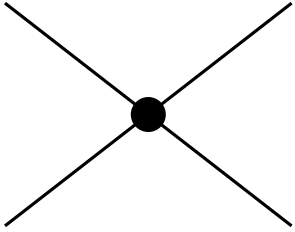}}
\caption{Leading order diagram for $\pi$-$\pi$ scattering.}
\label{f_pipi_0}
\end{figure}

As already mentioned, to order $D=6$ we have contributions from different
sources. First of all, form higher derivative terms in the Lagrangian and
secondly, from the expansion to higher order in the pion fields, giving rise to
loops. The beauty of chiral perturbation theory is, that the effects of loops
can be systematically be absorbed into renormalized couplings and masses. For
details see e.g \cite{DGH92}.

By now, the astute reader will have asked himself: How do I know, that a
momentum is small, or in other words, what is the expansion scale? 
There are several answers on the market. Georgi \cite{Geo84} argues, based
on renormalization arguments, that the scale should be $4 \pi f_\pi \sim 1 \,
\rm GeV$, whereas others argue \cite{Leu95b,DGH92},
that the mass of the lowest lying resonance  should give the scale, since this
is the energy, where the entire game seizes
 to work. This seems to be a reasonable
argument and, assuming that there is no $\sigma$-meson of mass $\sim 500 \, \rm
MeV$,  the mass of the $\rho$-meson should provide a reasonable benchmark.

So far we have worked in the chiral limit, i.e. assuming that the pion mass
vanishes. The explicit breaking of chiral symmetry is introduced by terms of
the form $\sim tr( U^+ + U)$ and 
and the simplest symmetry breaking is 
\be
\delta {\cal L}_{X \chi SB} = \frac{f_\pi^2 \, m_\pi^2}{4} tr( U^+ + U)
\simeq  4 - \frac12 m_\pi^2 \Phi^2 + {\cal O}(\Phi^4)
\ee
which to leading order in the pion-fields corresponds to a pion mass-term (the
constant term does not contribute to the dynamics). Again, one can
have many symmetry breaking term involving the above structure, such as
\be
tr( U^+ + U), \;\; tr (\partial_\mu U^+ \partial^\mu U)tr( U^+ + U) \ldots
\ee
so that an ordering scheme is necessary. Therefore, in the realistic case of
explicit chiral symmetry breaking, the scattering amplitudes are not only
expanded in terms of the pion momenta but also in terms of the pion
masses. The counting-rule is the same as given above \raf{count_pi}, where 
$d_i$ now gives the number of derivatives {\em and} pion masses of a given 
vertex of type
$i$. The total effective Lagrangian for pion-pion scattering 
to order $D=4$ is then given by
\be
{\cal L}_{2}^{(4)} = \frac12 (\partial_\mu \Phi)^2 - \frac12 m_\pi^2 \Phi^2 +
\frac{1}{6 f_\pi^2} \left[ (\vec{\Phi} \cdot \partial_\mu \vec{\Phi})^2 - 
\Phi^2 (\partial_\mu \vec{\Phi} \cdot \partial^\mu \vec{\Phi}) \right] +
\frac{m_\pi^2}{24 f_\pi^2} (\vec{\Phi} \cdot \vec{\Phi})^2 
\ee
In principle the `adjustable' parameters of this Lagrangian are the pion-mass
and the pion-decay constant, which have to be fixed to the 
experimental values.

The resulting  pion-pion scattering length and volumes are then given by
\cite{Wei66}
\be
a_0^0 = \frac{7 m_\pi}{32 \pi f_\pi^2}, \;\; 
a_0^2 = -\frac{ m_\pi}{16 \pi f_\pi^2}, \;\; 
a_1^1 = \frac{1}{24 \pi f_\pi^2 m_\pi }
\ee
where the subscript denotes the angular momentum and the superscript the
isospin of the amplitude.
As shown in table \raf{tab1} \cite{DGH92}, 
the leading order results agree reasonably well with
experiment and are improved by the next to leading order corrections.
Apparently we do not find perfect agreement with experiment even 
for the s-wave scattering lengths, although already to leading order 
we have taken into account terms quadratic in the momenta, 
so that higher orders in the pion momentum will not improve the situation.
However, remember, that 
we not only expand in terms of the pion momenta, but also, as a result of the
explicit symmetry breaking, in terms of the pion mass, which in principle
can contribute to any order to the s-wave scattering length. 

\begin{table}[t]
\begin{center}
\begin{tabular}{llll} \hline
& Experiment & Lowest Order & First Two Orders \\ \hline
$a_0^0 m_\pi$ & $  \;\; \: 0.26 \; \: \pm 0.05$ & $\;\; \: 0.16$ & 
$\;\; \: 0.20$ \\
$a_0^2 m_\pi$ & $-0.028 \pm 0.012$ & $-0.045$ & $ -0.041 $ \\
$a_1^1 m_\pi^3$ & $ \;\; \: 0.038 \pm 0.002$ & $\;\; \: 0.030$ &  
$ \;\; \:0.036 $ \\
\hline
\end{tabular}
\end{center}
\caption{Pion-pion scattering length}
\label{tab1}
\end{table} 

As already pointed out in the beginning of this section, chiral perturbation
theory, or more precisely, the expansion in momenta breaks down, once we get
close to a resonance. This one easily understands by looking at the
Breit-Wigner formula for the scattering amplitude involving a resonance.
\be
f(E) \sim  \frac{ \Gamma /2}{ E_R - E - i \Gamma / 2}
\ee
For energies, which are small compared to the resonance energy, $E \ll E_R$
this amplitude may be expanded in terms of a power series and the concept of
chiral perturbation theory works well
\be
f(E) \sim  \frac{ \Gamma /2}{ E_R } \left( 1 + \frac{E + i \Gamma / 2}{E_R} +
 \ldots\right);
\;\;\;  E \ll E_R
\ee

However, once we get close to the resonance-energy, we need to expand to higher
and higher order until at  $E \ge E_R$ the power-series in $E$ seizes to
converge. To be specific, we expect that in the the isovector p-wave channel,
which is dominated by the $\rho$-meson resonance, 
the chiral perturbation expansion should fail for energies $E \sim m_\rho$.

Finally, let us include the nucleons into the chiral counting. Naively, one
would think, that this should destroy the entire concept, because the nucleon
has a large mass, which is of the order of the expansion scale. However,
since at low energies the scattering amplitude may also be calculated in a
nonrelativistic framework, we do not expect the nucleon mass to enter directly,
but, to leading order,  only 
via the kinetic energy $\sim \frac{p^2}{2 M_N}$, 
which is small compared to that
of the pion at the same momentum. Therefore, chiral perturbation theory should
also work with nucleons present (for details see. \cite{Wei91}).
The above argument can be formalized by realizing that the nucleon only enters
the amplitudes through 
the nucleon propagator  (see e.g. the results of section
\raf{pn_scatter}).  At low momenta, the nucleon propagator  
contributing to diagram (a) of fig. \raf{diagram2}  can be written as
\be
\frac{\gamma_\mu (p^\mu + q^\mu) +M_N}{ (p + q)^2 - M_N^2} \simeq 
\frac{\gamma_0 M_N +M_N}{ 2 M_N q } = \frac{\Lambda}{q}
(1 + {\cal O}{(\frac{q}{M_N})} )
\label{nuc_prop}
\ee
where
\be
\Lambda =  \frac{\gamma_0 M_N +M_N}{2 M_N} = 
\left( \begin{array}{cc} 1 & 0\\0 & 0 \end{array} \right) 
\ee
projects on positive energy states. Hence, to leading order, each nucleon
propagator contributes like $\frac{1}{q}$ to the power of pion momentum of the
scattering amplitude. This leads to the following counting rule, which now also
includes the nucleons \cite{Wei91}
\be
D = 2 + 2 L - \frac{1}{2} E_N + \sum_i V_i(d_i + \frac12 n_i - 2)
\ee
Here the notation is as in equ. \raf{count_pi} and $E_N$ denotes the number of
external nucleon lines and $n_i$ the number of nucleon fields of vertex $i$,
which is typically $n_i = 2$.

For the simple nucleon-pole diagram using pseudovector coupling we thus would
have: $L = 0$, $E_N = 2$, $d = 1$, $n = 2$ such that, $d + \frac12 n - 2 = 0$
      and, $D = 1$.

On top of the expansion in terms of pion-momenta and pion masses, from
equ. \raf{nuc_prop} we, therefore, 
also have an expansion in the velocity of the
nucleons $v \sim \frac{q}{M_N}$. This is carried out in a systematic fashion in
the so called Heavy-Baryon Chiral-Perturbation Theory, as introduced by Jenkins
and Manohar \cite{JM92}. This approach essentially corresponds to a
systematic nonrelativistic expansion for the nucleon wave-function, on the
basis that the nucleon (baryon) is heavy compared to the momenta involved. 
We should mention, that the effect of the nucleon
can also be included in a fully covariant fashion as
discussed by Gasser et al. \cite{GSS88}.

Including the nucleon gives rise to additional structures which explicitly
break the chiral symmetry, such as
\be
\delta {\cal L} = a \, tr(U^+ + U) \bar{\psi} \psi 
\simeq a (1 - \frac{\phi^2}{2
f_\pi^2} ) \, \bar{\psi}\psi
\ee
To leading order, this is just a contribution to the nucleon mass, which allows
us to identify the coefficient $a$ with the sigma-term 
$\Sigma_{\pi N}$ (see section \ref{sec_3.2})
\be
\delta {\cal L} = - \Sigma_{\pi N} \, tr(U^+ + U) \bar{\psi} \psi \simeq 
- \Sigma_{\pi N} \, \bar{\psi}\psi  + 
\frac{\Sigma_{\pi N}}{2 f_\pi^2}\, \bar{\psi}\psi \,\phi^2
\ee
The next to leading term in the above expression is an {\em attractive}
interaction between pion and nucleon, which contributes to the order $D=2$ to
the amplitude. This term by itself is quite large  and would lead to
a wrong prediction for the s-wave pion-nucleon amplitude. However, there are
additional terms contributing to the same order, which in the heavy-fermion 
expansion come from the nucleon-pole diagrams. The coefficients of these terms
then need to be chosen such, that the resulting scattering length acquire the
small value observed in experiment \cite{TW95}.

\section{Applications}
In this last section, we want to discuss a few applications of chiral symmetry
relevant to the physics of dense and hot matter. First, we briefly address the
issue of in medium masses of pions and kaons. 
Then we will discuss the temperature and density dependence 
of the  quark condensate.
We will conclude with some general remarks on the properties of vector mesons 
in matter as well as on disoriented chiral condensates.

\subsection{Pion and kaon masses in dense matter}
Changes of the pion mass in the nuclear medium should show up in the 
iso-scalar pion s-wave optical potential. 
To leading order in the density this is related to the s-wave iso-scalar 
scattering-length $a_0^+$ by \cite{EW88}
\be
2 \omega U = - 4 \pi (1 + \frac{m_\pi}{M_N}) \, a_0^+ \, \rho
\label{opt_pot}
\ee
where $\omega$ is the pion energy. Since the s-wave iso-scalar scattering 
length
is small, as a result  of chiral symmetry, and slightly repulsive, we predict a
small increase of the pion mass in the nuclear medium, which at nuclear matter 
amounts to $\Delta m_\pi \simeq 5 MeV$. One arrives at the same result by
evaluating the effective Lagrangian, as obtained from chiral perturbation 
theory, at finite density \cite{BLR94,TW95}. 
This is not  surprising since the s-wave iso-scalar amplitude is
used to fix the relevant couplings.

In case of the kaons, which can also be understood as Goldstone bosons of an
extended $SU(3) \times SU(3)$ chiral symmetry, some interesting features
occur. Chiral perturbation theory predicts a repulsive s-wave scattering length
for $K^+$-nucleon scattering and a large  attractive one for $K^-$
\cite{KN86,BLR94}.  
Using the above relation for the optical potential \raf{opt_pot}
this led to speculations about a possible s-wave kaon condensate in dense
matter \cite{KN86,NK87} 
with rather interesting implications for the structure and stability of
neutron stars \cite{BTK92,TPL94}. 
Experimentally, however,  one finds that the  iso-scalar 
s-wave scattering length for the $K^-$ is repulsive, calling into question the
results from chiral perturbation theory. The resolution to this puzzle is the
presence of the $\Lambda (1405)$ resonance just below the kaon-nucleon
threshold. This resonance, which has not been taken into account in the chiral
perturbation analysis,  gives a large repulsive contribution to the scattering
amplitude at threshold. Does that mean, that chiral perturbation theory
failed? Yes and no. Yes, because, as already pointed out, it is not able to
generate any resonances and thus leads to bad  predictions in the neighborhood
of the resonance\footnote{Lee et al. \cite{LBR94} have attempted to include 
the $\Lambda (1405)$ as an explicit state in a chiral perturbation theory
analyses of the kaon-nucleon scattering length (see also \cite{Sav94}). 
While this approach 
may be a reasonable thing to do phenomenologically, it appears to be  
beyond the original  philosophy of chiral perturbation theory.}. 
No, because it predicts a strong attraction between the proton and the $K^-$, 
which, if iterated to infinite order can generate the $\Lambda
(1405)$-resonance as a bound state in the continuum \cite{KSW95}( in the
continuum, because the  $\Lambda (1405)$ decays into $\Sigma \pi$). 

This
situation is well known from nuclear physics. The proton-neutron scattering
length in the deuteron channel is repulsive although the proton-neutron
interaction is attractive. The reason is, that in this channel a
bound state can be formed, the deuteron, which gives rise to a strong repulsive
contribution to the scattering amplitude at threshold. 

To carry this analogy
further, we know that in nuclear matter the deuteron has disappeared,
essentially due to Pauli-blocking, revealing the true, attractive nature of the
nuclear interaction. As a result we have an attractive mean field
potential for the nucleons. Similarly, one can argue \cite{Koc94b}, that the 
$\Lambda (1405)$, if it is  a $K^-$-proton bound state, should
eventually disappear, resulting in an attractive s-wave optical potential for
the $K^-$ in nuclear matter. Indeed, an analysis of $K^-$ atoms \cite{FGB93}, 
shows, that the optical potential turns  attractive already at rather 
low densities $\rho \leq 0.5 \rho_0$. Extrapolated to nuclear matter density
the extracted optical potential would be as deep as $-200 \, \rm MeV$, in
reasonable agreement with the predictions from chiral perturbation theory.  
 
\subsection{Change of the quark-condensate in hot and dense matter}
\subsubsection{Temperature dependence}
\label{condensate}
One of the applications of chiral perturbation theory relevant to the physics
of hot and dense matter is the calculation of the temperature dependence of
the quark condensate. Here we just want derive the leading order
result. A detailed discussion, which includes also higher order corrections can
be found in ref. \cite{GL89}.
The basic idea is to realize that the operator of the quark-condensate, 
$\bar{q}q$, enters into the QCD-Lagrangian via the quark mass term. Thus, we
may write the QCD-Hamiltonian as
\be
H = H_0 + m_q \bar{q}q 
\label{h_hqcd}
\ee
The quark condensate at finite temperature  
is then given by the following statistical sum
\be
<\bar{q}q >_T = \frac{\sum_i <i| \bar{q}q \, e^{-H/T} |i>}
                   {\sum_i <i| e^{-H/T} |i> }
\ee
Since $\partial H / \partial m_q = \bar{q}q$ this can be written as 
\be
<\bar{q}q >_T = T \frac{\partial}{\partial m_q} \ln Z (m_q)
\ee
where the partition function $Z$ is given by $Z = \sum_i <i|e^{-H/T} |i>$.

In chiral perturbation theory we do not calculate the partition
function of QCD, but rather that of the effective Lagrangian. To make contact
with the above relations, we utilize the Gell-Mann Oakes Renner relation 
\raf{GOR}. To leading order in the pion mass the derivative with respect to the
quark mass, therefore, can be written as
\be
\frac{\partial}{\partial m_q} = - \frac{<0|\bar{q}q |0>}{f_\pi^2}
\frac{\partial}{\partial m_\pi^2}
\ee
Next to leading order contributions arise, among others, from the quark-mass
dependence of the vacuum condensate.

To leading order the partition function is simply given by that of a
noninteracting pion gas
\be
\ln \, Z =  \ln \, Z_0 + \ln \, Z_{\pi-\rm gas} =  \ln \, Z_0 + 
\frac{3}{(2 \pi)^3} \int d^3 p \,\ln (1 - \exp(-E/T))
\ee
where $Z_0$ stands for the vacuum contribution, which we, of course, cannot
calculate in chiral perturbation theory, since we are only concerned with
fluctuations around that vacuum.
Thus the temperature dependence of the quark condensate in the chiral limit is
given by
\be
<\bar{q}q >_T & = & \left. <0|\bar{q}q |0>  - \, T \, 
\frac{<0|\bar{q}q |0>}{f_\pi^2}
\frac{\partial}{\partial m_\pi^2} Z_{\pi-\rm gas} \right|_{m_\pi \rightarrow 0}
\nonumber \\
& = & <0|\bar{q}q |0> (1 - \frac{T^2}{8 f_\pi^2} )
\ee
Thus to leading order, the quark condensate drops like $\sim T^2$, 
i.e. at low temperatures the change in the condensate is small. 

Corrections include the effect of pion interactions, which in the
chiral limit  are proportional to the pion momentum 
and thus contribute to higher orders in the temperature. 
Including contributions  up to three loops, one finds (see e.g. \cite{GL89})
\be
\frac{<\bar{q}q >_T}{<\bar{q}q>_0}  =     1 - 
c_1 \left(\frac{T^2}{8 f_\pi^2} \right) - 
c_2  \left( \frac{T^2}{8 f_\pi^2} \right)^2 - 
c_3 \left(\frac{T^2}{8 f_\pi^2} \right)^3 \ln
  (\frac{\Lambda_q}{T} ) + {\cal O}(T^8) 
\ee
For $N_f$ flavors of massless quarks
the coefficients are given in the chiral limit by
\be
c_1 = \frac23 \frac{N_f^2 - 1}{N_f} \hspace{.8cm}
c_2 = \frac29 \frac{N_f^2 - 1}{N_f^2} \hspace{.8cm}
c_3 = \frac{8}{27}(N_f^2 + 1) N_f
\ee
The scale  $\Lambda_q$ can be fixed from pion
scattering data to be $\Lambda_q = 470 \pm 110\, \rm MeV$. 
In fig. \raf{qbq} we show the temperature dependence of the quark-condensate as
predicted by the above formula. Currently, lattice  gauge calculations predict
a critical temperature $T_c \simeq 150 \, \rm MeV$, above which the quark
condensate has disappeared. At this temperature chiral perturbation theory  
predicts only a drop of about 50~\%, which gets even smaller once pion masses
are included \cite{GL89}. However, we do not expect chiral
perturbation to work well close to the critical temperature. 
The strength of this approach is at low temperatures. The prediction, that
to leading order the condensate drops quadratic in the temperature is a
direct consequence of chiral symmetry and can be used to check chiral models as
well as any other conjectures involving the change of the quark-condensate, 
such as e.g. the change of hadron masses.

\begin{figure}[t]
\setlength{\epsfysize}{3in}
\centerline{\epsffile{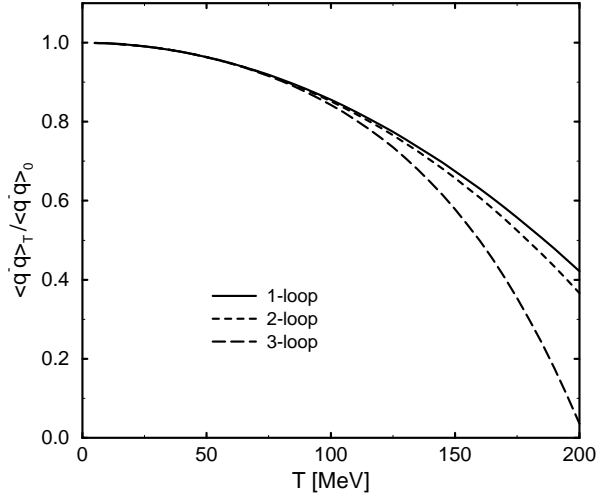}}
\caption{Temperature dependence of the quark condensate from chiral
perturbation theory (chiral limit).}
\label{qbq}
\end{figure}

\subsubsection{Density dependence}
For low densities, the density dependence of the quark condensate can also be
determined in a model independent way\footnote{Again, we give a heuristic
argument. A rigorous derivation based on the Hellmann-Feynman theorem can be
found e.g. in \cite{DL90,CFG92}.}. We expect that to leading order in
density the change in the quark
condensate  is simply given by the amount of quark
condensate in a nucleon multiplied by the nuclear density,
\be
<\bar{q}q>_\rho = <\bar{q}q>_0 + <N | \bar{q}q |N> \rho + \rm higher \; orders
\; in \; \rho 
\ee
All we need to know is the matrix element of $\bar{q}q$ between nucleon
states. This matrix element, however, enters into the pion-nucleon sigma-term,
eq. \raf{sig_pin2}
\be
<N | \bar{q}q |N> =  \frac{\Sigma_{\pi N}}{m_q} = - \Sigma_{\pi N} 
\frac{<\bar{q}q>_0}{m_\pi^2 f_\pi^2}
\ee
where we also have made use of the GOR-relation \raf{GOR}, namely 
$m_q = - \frac{m_\pi^2 f_\pi^2}{<\bar{q}q>_0}$. Thus we predict, that the quark
condensate drops {\em linearly} with density, as compared to the quadratic
temperature dependence found above
\be
<\bar{q}q>_\rho = <\bar{q}q>_0 (1 - 
 \frac{\Sigma_{\pi N}}{m_\pi^2 f_\pi^2} \rho + \ldots ) 
\ee
Corrections to higher order in density arrise, among others, 
from nuclear binding effects. These have been estimated 
\cite{LK94,BW95} to be at most of
the order of 15 \% for denities up to twice nuclear-matter density.
Assuming a value for the sigma term of $\Sigma_{\pi N} \simeq 45 \, \rm MeV$ we
find that the condensate has dropped by about 35 \% at nuclear matter density
\be
<\bar{q}q>_\rho = <\bar{q}q>_0 (1 - 0.35 \frac{\rho}{\rho_0})
\ee

Thus, finite density is very efficient in reducing the quark condensate and we
should expect that any in medium modification due to a dropping quark
condensate should already be observable at nuclear matter density.
The above findings  also suggest, that chiral restoration, i.e. the vanishing
of the quark-condensate, is best achieved in heavy ion collisions at bombarding
energies, which still lead to full stopping of the nuclei.

\subsection{Masses of vector mesons}
Finally, let us briefly discuss what chiral symmetry tells us about the masses
of vector mesons in the medium. Vector mesons, such as the $\rho$-meson, are of
particular interest, because they decay into dileptons. Therefore, possible
changes of their masses in medium are accessible to experiment. 

Using current algebra and PCAC, Dey et al. \cite{DEI90} could show,
that at finite temperature the mass of the rho-meson does not change to 
order $T^2$. Instead to order $T^2$ 
the vector-correlation function gets an admixture from the axial-vector 
correlation function
\be
C_V(T) = (1 - \epsilon) \, C_V(T=0) + \epsilon \, C_A(T=0) 
\label{dey_spec}
\ee
with $\epsilon =\frac{T^2}{ 6 f_\pi^2 }$.  
The imaginary part of this vector-correlation function is 
directly related to the dilepton-production cross-section. 
As depicted in fig. \raf{rhomass}, the above result, therefore, predicts 
that to leading order in the temperature, the dilepton invariant mass spectrum 
develops a peak at the mass of the $a_1$-meson in addition to that
at the mass of the $\rho$.
At the same time, the contribution at the $\rho$-peak is
reduced in comparison to the free case. Furthermore, the position of the peaks
is not changed to this order in temperature. This general result is also
confirmed by calculations in chiral models, which have been extended to include
vector mesons \cite{Son93,Pis95}. Notice, that the above finding also rules
out that the mass of the $\rho$-meson scales linearly with the quark
condensate, because previously (see section \ref{condensate}) we found that the
quark condensate already  drops to order $T^2$, whereas the mass of the
$\rho$ does not change to this order. 

\begin{figure}[t]
\setlength{\epsfxsize}{4in}
\centerline{\epsffile{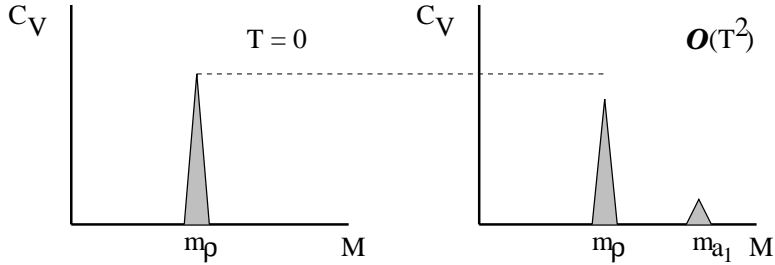}}
\caption{Vector-spectral functions at $T=0$ and to leading order in
temperature as given by equ. \protect\raf{dey_spec}.}
\label{rhomass}
\end{figure}

Corrections to higher order in the temperature, however, are not controlled by
chiral symmetry alone and, therefore, one finds model dependencies. Pisarski
\cite{Pis95} for instance predicts in the framework of a linear sigma model
with vector mesons, that to order $T^4$ the mass of the $\rho$ decreases and
that of the $a_1$ increases. Song \cite{Son93,Son96}, 
on the other, uses a nonlinear
$\sigma$-model and finds a slight increase of the $\rho$-mass 
as well as a dropping of the $a_1$-mass. 
At the critical temperature, both again agree qualitatively
in that the masses of $a_1$ and $\rho$ become degenerate at a value which is
roughly given by the average of the vacuum masses $\simeq 1 \, GeV$. 
This agreement, again, is a result of chiral symmetry.
  
At and above the critical temperature, where chiral symmetry is not anymore 
spontaneously broken, chiral symmetry demands that the 
vector and axial vector correlation functions are the same. One way to 
realize that is by the having the same masses for the vector ($\rho$) and
axial-vector ($a_1$). 
However, this is not the only possibility! As nicely discussed in a paper
by Kapusta and Shuryak \cite{KS94}, there are at least three
qualitatively different possibilities, which are sketched in
fig. \raf{restore}. 
\begin{enumerate}
\item
The masses of $\rho$ and $a_1$ are the same. In this case, clearly the vector
and axial vector correlation functions are the same. Note, however, that we
cannot make any statement about the value of the common mass. It may be zero,
as suggested by some people, it may be somewhere in between the vacuum 
masses, as the chiral models seem to predict and it my be even much larger 
than the mass of the $a_1$.
\item
We may have a complete mixing of the spectral functions. Thus, both 
the vector and axial-vector spectral functions have peaks of equal
strength at both the mass of the $\rho$ and the mass of the $a_1$, leading to
two peaks of equal strength in the dilepton spectrum (modulo Boltzmann-factors
of course). One example would be given by the low temperature result
\raf{dey_spec} with 
the mixing parameter $\epsilon(T_c) = \frac12$. 
Using the low temperature result for 
$\epsilon = \frac{T^2}{ 6 f_\pi^2 }$ would give a critical temperature
of $T_c = \sqrt{3} f_\pi \simeq 164 \, MeV$, which is surprisingly close to
the value given by recent lattice calculations.
\item
Both spectral functions could be smeared over the entire mass range. Due to 
thermal broadening of the mesons and the onset of deconfinement, the structure
of the spectral function may be washed out and it becomes meaningless to talk
about mesonic states.
\end{enumerate}

\begin{figure}[t]
\setlength{\epsfxsize}{5.5in}
\centerline{\epsffile{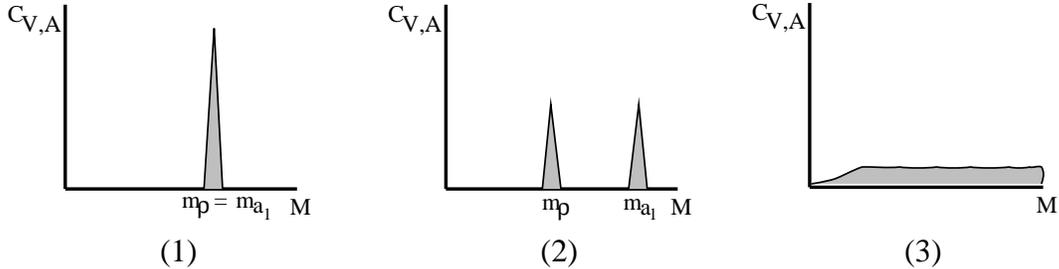}}
\caption{Several possibilities for the vector and axial-vector spectral
functions in the chirally restored phase.}
\label{restore}
\end{figure}

To summarize, the only unique prediction derived from chiral symmetry (current
algebra) about the
temperature dependence of the $\rho$-mass, is that it does not change to order
$T^2$, i.e. at low temperatures. Furthermore, at and above the critical
temperature, chiral symmetry requires that the vector and axial-vector spectral
functions are identical, which, however, does not necessarily imply, that both
exhibit just one peak, located at the same position.
Corrections of the order $T^4$ cannot be obtained from chiral symmetry alone.

Finally let us point out, that the above findings do not rule out scenarios,
which relate the mass of the $\rho$ with the temperature dependence of the 
bag-constant or gluon condensate, such as proposed by Pisarski
\cite{Pis82} and Brown and Rho \cite{BR91}. These ideas, however, involve
concepts which go beyond chiral symmetry, such as the melting of the gluon
condensate. Consequently in these scenarios, a certain behavior of the mass 
of the $\rho$-meson
can only indirectly be brought in connection with chiral restoration.

\subsection{Disoriented Chiral Condensates (DCC)}
One interesting aspect of the restoration of chiral symmetry in relativistic
heavy ion collision is the possible formation of a so called disoriented chiral
condensate (DCC). The idea of the formation of such a DCC was first put forward
by Anselm \cite{Ans89}, Bjorken \cite{Bjo92} (he called it `Baked Alaska'),
Blaizot and  Krzywcki \cite{BK92}, and 
by Rajagopal and Wilczek \cite{RW93}, and it 
has sparked a great deal of theoretical activity. The
status of this field is nicely reviewed in \cite{Raj95} and we refer the
reader to this reference for all the details which we will not present
here.
 
What is a DCC? Simply speaking, it is a lump
of possibly coherent pions of the same type. 
Thus, contrary a gas of free pions, where everywhere in space, 
one has the same number 
of $\pi^0$, $\pi^+$, and $\pi^-$, in case of DCC formation,  pions of the same 
charge state  are lumped together. Now, if space and rapidity are correlated, 
as one expects to be the case in ultrarelativistic heavy ion collisions, DCC
formation should lead to large fluctuations of the ratio of, say, 
$\pi_0 / (\pi^+ + \pi^- + \pi^0)$, 
as a function of rapidity. Furthermore, if the pions are lumped
together over a volume of reasonably large size, 
the typical momentum of these pions
is of the order of $1/R$, where $R$ is the radius of the `lump'. If $R \geq 2
\, \rm fm$, the typical momentum is much smaller than the thermal one, which
could lead to another observable, namely an enhancement in the low momentum 
part of the pions spectrum\footnote{This will only work, however, if the pions
from the DCC do not reinteract with the thermal pions surrounding them. To
which extent this is the case in a relativistic heavy ion collision, remains 
to be seen.}.

Now, how does one form a DCC and what does it have to do with chiral symmetry
restoration?

First, let us rule out the explanation which might come to mind first.
If, as is currently believed,  the chiral
phase transition is of second order in the limit of 2 massless quarks, 
the fluctuations of the order parameter (namely the pion and the sigma fields)
should generate long range correlations. Naively, one, therefore, would 
expect that these long range fluctuations could generate large domains 
of the pion fields. However, the problem is, that we do not have 
massless quarks, i.e. that
chiral symmetry is explicitly broken. As a result the pion has a mass, which
actually slightly increases as one approaches the critical point (see e.g.
ref. \cite{Raj95}). As a consequence, the range of the fluctuations of the 
order parameter is restricted to the inverse of the pion mass, 
about $1.4 \, \rm fm$, and we do not expect domains of 
large size to be created.

\begin{figure}[thb]
\setlength{\epsfysize}{2.5in}
\centerline{\epsffile{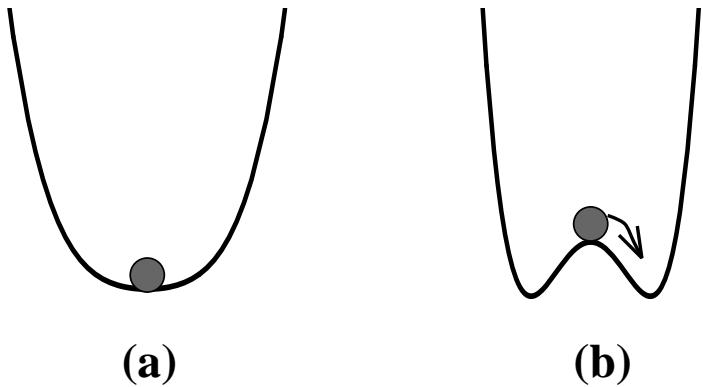}}
\caption{System in restored phase (a) and in quench scenario (b).}
\label{fig11}
\end{figure}

Domains of considerable size, however, can be formed by a dynamical
process. Assume chiral symmetry is restored initially. Thus, the expectation
values of the $\sigma$ as well as pion fields are zero and the effective
Hamiltonian has the shape depicted in figure \ref{fig11} (a) (for a three
dimensional view see figure \ref{sponti} (a)).  If the system
undergoes a so called quench \cite{RW93}, meaning that the state is still
in the center of the effective potential, but the effective potential has
changed to
that of the zero temperature state, the state find itself all of a sudden 
at a maximum of the effective potential. This situation is shown in figure 
\ref{fig11}(b). This situation is unstable, and the instabilities
associated with that generate an enhancement of the low momentum modes, and,
hence, long range correlations.

In order to see, that only the low momentum modes are enhanced, lets consider
the equation of motion for the pion as derived from the Lagrangian of the
linear sigma model (eq. \raf{L_LS}) without nucleons.

\be
\frac{\partial^2}{\partial t^2} \phi - \nabla^2 \phi + 
\lambda ( (\phi^2 + \sigma^2) - v_0^2) \phi = 0
\ee
where $v_0^2 = f_\pi^2 - \frac{m_\pi^2}{\lambda}$ (see eq. \raf{eq:v_0})
as a result of the explicit symmetry breaking. 
Going to momentum space and introducing an effective mass, $m_{eff}(k,t)$,
this equation can be approximately rewritten as
\be
\frac{\partial^2}{\partial t^2} \phi(k,t) = - (k^2 + m_{eff}^2(t)) \phi(k,t)
\ee
with 
\be
m_{eff}^2(t) = \lambda ( \langle \phi^2 + \sigma^2 \rangle - v_0^2)  
\ee
At zero temperature, in the
ground state, $\langle \sigma^2 \rangle= f_\pi^2$, $ \langle \Phi \rangle =
0$, and we obtain the equation of motion for the
pion with mass $m_\pi$. In the quench scenario (figure \ref{fig11}(b)), 
however,
the expectation value of the sigma field vanishes and square of the effective 
mass becomes negative leading to an amplification of the modes with small
mometum $k$!
More precisely, since at finite temperatures thermal fluctuations contribute to
the expectation value of $\phi^2$ and $\sigma^2$, in the quench scenario we
have the following effective mass
\be
m_{eff}^2(t) = \lambda ( \langle \phi^2 + \sigma^2\rangle_{thermal} - v_0^2)
\ee
If $\langle \phi^2 + \sigma^2 \rangle_{thermal} \, < v_0^2$,
the square of the effective mass becomes negative. Consequently  modes with 
momenta  $k^2 < |m_{eff}^2|$ become
exponentially amplified whereas modes with higher momentum are not
enhanced. Since it is the low momentum modes that become amplified, long range
correlations start to build up, leading to what is called a DCC.
(For a detailed discussion including numerical results we refer the reader to
\cite{Raj95}.)

How realistic is the quench scenario? This question is actually still
debated. Numerical studies based on the linear sigma model show (see e.g. 
\cite{AHW95,Ran96}), that a rapid expansion of the system, as one expects it
to take place in heavy ion collisions, can drive the system into the unstable
region. These calculations, however, are based on only four degrees of freedom
whereas lattice gauge calculations show that the entropy at the chiral
transition point corresponds to that of about 38 degrees of freedom. With a
considerably larger initial entropy density it remains to be seen if the
expansion is sufficient to cool the system fast enough in order to drive 
it in into the unstable region.

Finally, there is the question of observation. Clearly if the creation and
existence of a DCC could be confirmed in experiment, this would be a rather
unambiguous signal that the system has crossed the phase transition line. There
are many suggestions on how to observe a DCC state, ranging from pion spectra
\cite{Gav95}
to fluctuations in the ratios of the pion charge states \cite{Raj95}. 
Electromagnetic signals seem to be particularily promising \cite{KKR96}. 
However, at the
present stage, all these suggestions need to be put into realistic model
calculations for heavy ion collisions in order to see, if these signal are not
altered or overshadowed by the hadronic environment created in such a
collision. 

\ \\
\ \\
\noindent
{\bf Acknowledgments:} I would like to thank C. Song for useful discussions
concerning the effects of chiral symmetry on the vector mesons. 
This work was supported by the Director, 
Office of Energy Research, Office of
High Energy and Nuclear Physics Division of the Department of Energy, under
Contract No. DE-AC03-76SF00098.

\end{document}